\documentclass[12pt]{iopart}

\usepackage{iopams}  
\usepackage{bbm}
\usepackage{slashed}
\usepackage{graphicx}
\usepackage[breaklinks=true,colorlinks=true,linkcolor=blue,urlcolor=blue,citecolor=blue]{hyperref}
\usepackage[english]{babel}
\usepackage[usenames]{color}
\usepackage{cite}
\usepackage{tikz}

\renewcommand{\Re}{{\rm Re}}

\newcommand{\Db}{{\bar{D}}}
\newcommand{\Delb}{{\bar{\Delta}}}
\newcommand{\gb}{{\bar{g}}}
\newcommand{\Rb}{{\bar{R}}}

\newcommand{\cL}{{\mathcal L}}
\newcommand{\cM}{{\mathcal M}}
\newcommand{\cN}{{\mathcal N}}
\newcommand{\cO}{{\mathcal O}}
\newcommand{\cR}{{\mathcal R}}
\newcommand{\cT}{{\mathcal T}}

\newcommand{\p}{{\partial}}

\newcommand{\unit}{\mathbbm{1}}

\newcommand{\be}{\begin{equation}}
\newcommand{\ee}{\end{equation}}
\newcommand{\ba}{\begin{eqnarray}}
\newcommand{\ea}{\end{eqnarray}}

\newcommand{\addsto}{\ensuremath{\stackrel{+}{\longmapsto}}}

\begin{document}

\title[Form Factors in Asymptotic Safety]{Form Factors in Asymptotic Safety: \\ conceptual ideas and computational toolbox}

\author{Benjamin Knorr$^1$\href{https://orcid.org/0000-0001-6700-6501}{\includegraphics[scale=.07]{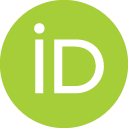}}, Chris Ripken$^1$\href{https://orcid.org/0000-0003-2545-5047}{\includegraphics[scale=.07]{ORCIDiD_icon128x128.png}} and Frank Saueressig$^1$\href{https://orcid.org/0000-0002-2492-8271}{\includegraphics[scale=.07]{ORCIDiD_icon128x128.png}}}

\address{$^1$ Institute for Mathematics, Astrophysics and Particle Physics (IMAPP),\\
	Radboud University, Heyendaalseweg 135, 6525 AJ Nijmegen, The Netherlands}

\eads{\mailto{b.knorr@science.ru.nl}, \mailto{a.ripken@science.ru.nl}, \mailto{f.saueressig@science.ru.nl}}

\vspace{10pt}
\begin{indented}
\item[] \date{\today}
\end{indented}

\begin{abstract}
Over the last years the Asymptotic Safety program has matured into a serious candidate for a quantum theory of gravity compatible with observations. The rapid technical progress in computing renormalisation group flows for gravity and gravity-matter systems in the non-perturbative regime has put many interesting physical questions within reach. In particular, the construction of the non-perturbative quantum corrections to the propagation of fields on a fluctuating spacetime allows addressing the effective propagation of matter on a quantum spacetime or the possible resolution of spacetime singularities based on first principle computations. In this article, we assemble a technical toolbox for carrying out investigations on this promising research frontier. As a specific example we present results for the momentum-dependent two-point function for a scalar field induced by the quantum fluctuations of the underlying geometry in a self-consistent way.
\end{abstract}

\vspace{2pc}

\noindent
\submitto{\CQG}
\maketitle
% 

%--------------------------------------------------------------------------
\section{Introduction}
\label{sec:intro}
%--------------------------------------------------------------------------
Relativistic quantum field theories have provided some of the most accurate predictions for physics experiments to date. In particular, the Standard Model of particle physics, underlying the data analysis at the Large Hadron Collider, provides a remarkably accurate account on how nature works up to energy scales of approximately 10 TeV. A rather intriguing prediction arising from the quantum field theory framework is that couplings, describing the strength of interactions between fundamental particles, depend on the energy of the process. They are promoted to scale-dependent or ``running'' quantities. A prototypical example for this behaviour is provided by the fine-structure constant $\alpha$, whose value at vanishing momentum is approximately $\alpha = 1/137$, while at the mass of the $Z$-boson it has already increased to $\alpha = 1/128$ \cite{Tanabashi:2018oca}.

A conceptual deficit of the Standard Model of particle physics is that gravitational interactions are not included. In fact, applying the perturbative quantisation techniques featuring in particle physics to Einstein's theory of General Relativity results in a quantum field theory which is perturbatively non-renormalisable. The negative mass dimension of Newton's coupling introduces new divergences at every loop order \cite{'tHooft:1974bx}. Since each divergence requires introducing a new, undetermined parameter, this essentially turns gravity into an effective field theory where the Einstein-Hilbert action provides the leading terms in the low-energy description. Owed to the exploding number of free parameters linked to physics at the Planck scale, combined with the notorious difficulties to measure them experimentally \cite{Aghanim:2018eyx, Ghosh:2019twk}, this framework is unsatisfactory when trying to address questions related to the structure of spacetime at Planckian resolution, the propagation of particles in regimes where fluctuations of spacetime cannot be neglected, or the fate of classical spacetime singularities within the quantum theory.

The Asymptotic Safety mechanism \cite{Weinberg:1976xy,Weinberg:1980gg} may improve this situation by 
providing a consistent and predictive theory of quantum gravity within the arena of relativistic quantum field theories. The underlying key idea, reviewed in e.g.\ \cite{Niedermaier:2006wt,Reuter:2012id,Percacci:2017fkn,Eichhorn:2018yfc,Reuter:2019byg}, is that gravitational interactions at trans-Planckian energies are controlled by a non-Gaussian renormalisation group fixed point (NGFP). At such a fixed point all dimensionless couplings remain finite, thereby removing the unphysical divergences appearing in physical processes. Moreover, the condition that the fixed point controls physics in the trans-Planckian regime places constraints on the free parameters appearing in the effective field theory approach. This enhanced predictive power may provide interesting predictions for the propagation of particles in a fluctuating spacetime, the physics related to event horizons, or the fate of spacetime singularities 
based on first principles computations. Quite remarkably, the asymptotic safety mechanism may also be operative when gravity is supplemented by suitable sets of matter fields \cite{Dona:2013qba}, which ultimately may provide a framework for unifying all fundamental forces. 

The primary tool for investigating the existence and properties of suitable NGFPs in gravity and gravity-matter systems is the functional renormalisation group (FRG), foremost the Wetterich equation \cite{Wetterich:1992yh} for the effective average action $\Gamma_k$. In this approach, one typically identifies projections of a fixed point by computing the RG flow on a subspace of all possible action functionals. This entails that the ``fundamental action'' associated with the theory is not known a priori but needs to be constructed from the fixed point data \cite{Manrique:2008zw,Manrique:2009tj,Morris:2015oca}. Moreover, one has to check that a candidate NGFP 
 meets general expectations concerning, e.g., the stability of the vacuum, degrees of freedom associated with the gravitational theory, or unitarity requirements. 
By now many of these questions have been investigated by studying projections of the NGFP 
onto spaces spanned by a finite number of interaction monomials. Typically, the interactions included in the projection subspace are selected based on power-counting arguments and computational feasibility. A prototypical example is provided by the polynomial expansion of $f(R)$-gravity \cite{Machado:2007ea,Codello:2007bd,Benedetti:2012dx,Falls:2014tra,Ohta:2015efa,Eichhorn:2015bna,Falls:2016wsa,Alkofer:2018fxj,deBrito:2018jxt}, where  $R$ is the Ricci scalar, or expansion schemes where all terms up to a certain number of spacetime-derivatives are retained in $\Gamma_k$ (derivative expansion) \cite{Benedetti:2009gn,Benedetti:2009rx,Benedetti:2010nr,Groh:2011vn,Hamada:2017rvn}.

It is clear, however, that these projection schemes are unreliable when addressing questions related to the propagation of fields and degrees of freedom supported by the NGFP. Performing a derivative expansion of terms contributing to the propagators and truncating the expansion at a finite order will almost certainly lead to results that are in conflict with Ostrogradski stability \cite{Becker:2017tcx,Arici:2017whq}. Finding reliable answers to these questions then requires tracing the renormalisation group flow of functions, the so-called form factors, which besides a dependence on the coarse-graining scale $k$ also depend on the momenta of the fluctuation fields. For gravity, the relevant terms are
\ba\label{structure1a}
\Gamma_k^{\rm C} & =  \frac{1}{16 \pi G_k}  \int \rmd^4x \sqrt{g} \, C_{\mu\nu\rho\sigma} \, W_k^{\rm C}(\Delta) \, C ^{\mu\nu\rho\sigma} \, , \\ \label{structure1b}
\Gamma_k^{\rm R} & =  \frac{1}{16 \pi G_k} \int \rmd^4x \sqrt{g} \, R \, W_k^{\rm R}(\Delta) \, R \, , 
\ea
where $C_{\mu\nu\rho\sigma}$ denotes the Weyl tensor, see \eref{Weyldef}, and $\Delta \equiv - g^{\mu\nu} D_\mu D_\nu$ is the Laplacian constructed from the metric $g_{\mu\nu}$. When expanded around a (Euclidean) flat space background the form factors $W_k^{\rm C}(\Delta)$ and $W_k^{\rm R}(\Delta)$ provide non-trivial contributions to the propagators of the transverse-traceless and  trace-part of the gravitational fluctuations. Moreover, the form factors $W_k^{\rm C}(\Delta)$ and $W_k^{\rm R}(\Delta)$ together with their generalizations related to terms containing more than two powers of the curvature tensors encode the momentum dependence of the ``running coupling constants'' at the level of the effective action $\Gamma_{k=0}$. Owed to their prominent role, e.g., in establishing the quantum propagators of the theory, knowing the form factors will be essential for understanding the spacetime structure entailed by Asymptotic Safety.

A first step in this direction was taken in \cite{Christiansen:2014raa}, where it was shown that 
the non-trivial momentum dependence of the diagrams  entering the flow equation (cf. \fref{fig:feynman}) generates a non-trivial momentum dependence of the form factors, even if one starts from quantising the Einstein-Hilbert action in a flat space background. The resulting induced form factors show that non-perturbative effects play an important role in the momentum dependence of the graviton propagator. In a related study \cite{Bosma:2019aiu}, but in a curved background, one of the form factors has been calculated in a self-consistent way for the first time with the help of the non-local heat kernel \cite{Codello:2012kq}. In \cite{Christiansen:2015rva, Denz:2016qks}, also the three- and four-point function were resolved, giving some information on these and higher-order form factors.

The goal of this compendium is to discuss the computation of momentum-dependent form factors from first principles in a self-consistent way. At the technical level, this comprises three pillars. Firstly, we provide a collection of analytic identities which allows to expand terms of the form \eref{structure1a} and \eref{structure1b} in powers of the fluctuation fields \emph{without making approximations on the momentum dependence of the resulting interaction vertices}. Secondly, we construct a prototypical example of a non-linear integro-differential equations capturing the scale- and momentum-dependence of the form factors. Thirdly, we introduce a set of numerical techniques for finding non-perturbative approximate solutions to these equations using  pseudo-spectral methods \cite{Boyd:ChebyFourier}. We expect that this well-stocked toolbox will be a solid foundation for computing the momentum-dependent propagators and vertices in Asymptotic Safety in the upcoming years.

The rest of the work is organised as follows. \Sref{sec:2.1} introduces the Wetterich equation and its essential properties while \sref{sec:FRG-qg} introduces two commonly used projection strategies for organising approximate solutions to the flow equation. The form factors determining the propagation of graviton and matter fluctuations are discussed in \sref{sec:struc-funcs}. This includes a detailed discussion of split-symmetry restoration in \sref{sec:struc-funcs-scal-tens}. \Sref{sec:scal-tens} contains the derivation of the integro-differential equation governing the scale-dependence of the form factor for scalar matter fields introduced in \eref{Sscalar}. The result \eref{structuremaster} serves as a prototypical example exhibiting all the structures expected for the flow equations in the other gravity-matter sectors. The properties and fixed function solution of the equation is then constructed in \sref{sec:scal-tens-fluc-st} and we comment on the properties of the solution in the conclusion, \sref{sec:discussion}.
Our notation and conventions are gathered in \ref{app:notation}. Technical details are provided in \ref{app:expansion} and \ref{app:laplacian}, while the rather lengthy computation related to the scalar-graviton four-point vertex has been relegated to \ref{app:4ptvertex}. Two Mathematica notebooks illustrating the derivation of the flow equation ({\tt flowequation.nb}) and the construction of the numerical solution ({\tt fpsolver.nb}) are provided as supplementary material.

%------------------------------------------------------------------
\section{The Functional Renormalisation Group Equation}
\label{sec:FRG}
%------------------------------------------------------------------
We start by introducing the main tool for investigating Asymptotic Safety, the functional renormalisation group equation (FRGE), governing the scale-dependence of the effective average action $\Gamma_k$. Upon highlighting some structural properties of the flow equation, we discuss the two main approaches towards constructing non-perturbative approximate solutions of this equation. Our main focus will be on the form factors describing the propagation of fluctuation fields in the gravitational and matter sectors.
%------------------------------------------------------------------
\subsection{The Wetterich equation}
\label{sec:2.1}
%------------------------------------------------------------------
The primary technical tool for investigating the existence and properties of renormalisation group fixed points suitable for Asymptotic Safety is the  Wetterich equation for the effective average action $\Gamma_k$ \cite{Wetterich:1992yh},
\be\label{FRGE}
\p_t \Gamma_k = \frac{1}{2} {\rm Tr}\left[ \left(\Gamma_k^{(2)} + \cR_k \right)^{-1}  \, \p_t \cR_k \right] \, ,
\ee
adapted to gravity first in \cite{Reuter:1996cp}. Here $t \equiv \log k$ is the RG time, $\Gamma_k^{(2)}$ denotes the second functional derivative of $\Gamma_k$ with respect to the fluctuation fields, and Tr contains a sum over all fields and indices as well as a functional trace. The operator $\cR_k$ provides an infrared regularisation of modes with momenta $p^2 \lesssim k^2$. The factor $\p_t \cR_k$ provides an ultraviolet (UV) regularisation, essentially ensuring that the high-energy modes with $p^2 \gtrsim k^2$ do not contribute to the trace. As a result, the change of $\Gamma_k$ is driven by  integrating out quantum fluctuations with momenta $p \simeq [k-\delta k, k]$, thereby realising Wilson's idea of renormalisation.
The FRGE \eref{FRGE} can be derived on rather generic grounds. Any ``theory space'' $\cT$, specified by fixing the field content entering into $\Gamma_k$ together with the underlying symmetries, will support a flow equation of the form \eref{FRGE}. The high degree of flexibility then allows to investigate gravity and gravity-matter systems within the same technical framework.  

Solutions of the FRGE, so-called RG trajectories, describe the same physical system at different coarse-graining scales $k$. They interpolate between the initial condition $\Gamma_{k = \Lambda_{\rm UV}} = S$ imposed at the UV-scale $\Lambda_{\rm UV}$ and the effective action $\Gamma_{k =0} = \Gamma$ generating the one-particle irreducible correlation functions of the full quantum theory, see \cite{Berges:2002ew,Pawlowski:2005xe,Gies:2006wv,Delamotte:2007pf,Nagy:2012ef,Percacci:2017fkn,Reuter:2019byg} for reviews. Notably, it is not necessary to introduce a ``fundamental action'' a priori. Suitable renormalisation group fixed points appear as properties of the renormalisation group flow and can often be seen in already simple approximations. 

A central advantage of the FRGE is that the functional nature of the equation allows to construct approximate (projected) solutions of the formally exact equation without the need to expand in a small coupling constant. The construction starts by introducing a set of interaction monomials $\{\cO_i \}$ spanning a basis on $\cT$. By definition the $\cO_i$ are constructed from the field content and obey the symmetries of the underlying theory space. Typical examples built from the spacetime metric $g_{\mu\nu}$ are the interaction monomials of the Einstein-Hilbert action (including a cosmological constant term),
\ba\label{O01}
\cO_\Lambda = & \int \rmd^dx \sqrt{g} \, , \qquad 
\cO_{\rm R} =  \int \rmd^dx \sqrt{g} R \, .
\ea
Here $R$ is the Ricci scalar constructed from $g_{\mu\nu}$. The effective average action can then be expanded in this basis,
\be\label{gammaexact}
\Gamma_k = \sum_i \bar{u}_i(k) \, \cO_{i} \, ,
\ee
where the expansion coefficients are given by scale-dependent couplings. Substituting the expansion \eref{gammaexact} into the FRGE \eref{FRGE} and equating the coefficients multiplying the basis elements $\cO_{i}$ yields the beta functions 
\be\label{betadimful}
\p_t \bar{u}_i(k) = \bar \beta_i(\{ \bar{u}_i\}, k) \, ,
\ee
capturing the scale dependence of the expansion coefficients. The explicit $k$-dependence can be removed by introducing dimensionless coupling constants
\be\label{udimless}
u_i \equiv k^{-d_i} \, \bar{u}_i \, ,
\ee
where $d_i$ is the mass dimension of the dimensionful coupling $\bar{u}_i$. Rewriting \eref{betadimful} in terms of the dimensionless couplings yields the dimensionless beta functions
\be\label{dimless}
\p_t u_i(k)  = \beta_i(u_1,u_2,u_3, \cdots) \, .
\ee
The $\beta_i$ constitute an autonomous vector field on theory space $\cT$ whose integral curves connect the same physics at different scales. RG fixed points then correspond to points $\{u_i^\ast\}$ where all beta functions vanish simultaneously,
\be
\beta_i(u_1^\ast,u_2^\ast,u_3^\ast, \cdots) = 0 \, , \quad \forall i \, . 
\ee

In a practical computation, the expansion \eref{gammaexact} is typically truncated to a finite subset of the $\cO_i$, $i=1,\cdots,N$. The general procedure then results in a coupled set of $N$ non-linear ordinary differential equations (ODEs) which can be integrated numerically. The search for fixed points reduces the system to finding the roots of a set of algebraic equations. Once a fixed point is identified, the RG flow in its vicinity can be studied by linearising the beta functions around $u_i^\ast$,
\begin{equation}
\beta_{u_i}( u_1, u_2, u_3, \cdots)
\simeq
\sum_j \mathbf{M}_{ij}	(u_j - u_j^\ast)
\, . 
\end{equation}
The matrix $\mathbf{M}_{ij} = \partial \beta_{u_i} / \partial{u_j}$ denotes the stability matrix.  Diagonalising the stability matrix allows us to write down the solution to the linearised flow in terms of the eigenvectors $V_I$ and eigenvalues $\theta_I$ of $\mathbf{M}$, 
\begin{equation}
u_i(t)
=
u_i^\ast
+	C_I	\, V_{Ii} \,	\exp(-\theta_I t)
\, , 
\end{equation}
where the $V_I, \theta_I$ are obtained from solving the eigenvalue problem $\mathbf{M} \, V_I = - \theta_I V_I$, and no sum over $I$ is implied. The integration constants $C_I$ determine the initial conditions of the flow. For stability coefficients with negative real part, $\Re(\theta_I) < 0$, the only way to end at the fixed point as $t\to \infty$ is if the corresponding $C_I$ vanishes. In this case, the eigendirection $V_I$ is called UV-irrelevant. Conversely, directions connected to positive critical exponents automatically run into the fixed point. The corresponding integration constants $C_I$ are undetermined by the Asymptotic Safety condition and parameterise the freedom of constructing asymptotically safe RG trajectories.

In special cases, the set of the $\cO_i$ may be extended to include an infinite number of interaction monomials. The prototypical example is given by the $f(R)$-truncation \cite{Codello:2007bd,Machado:2007ea,Benedetti:2012dx,Dietz:2012ic,Demmel:2014hla,Demmel:2015oqa,Morris:2016spn,Christiansen:2017bsy,Falls:2017lst,Alkofer:2018baq,deBrito:2018jxt} where $\Gamma_k$ contains an arbitrary, scale-dependent function $f$ of the Ricci scalar,
\be
\Gamma_k^{f(R)} = \int \rmd^dx \sqrt{g} \, f(R) \, . 
\ee
More generally, such a coupling function can depend on several arguments.
Evaluating the FRGE for this ansatz yields a non-linear partial differential equation (PDE) governing the scale-dependence of the function $f(R)$. In this setting, a fixed point then corresponds to a $k$-stationary solution of the corresponding dimensionless equation. The ODE encoding the fixed point solution may then either be used as a generating functional for describing a fixed point in a small-$R/k^2$ expansion or solved for global solutions numerically. 

The discussion on the stability extends straightforwardly to the case of a set of functions of field operators. The fixed point condition in general is then a set of PDEs with one variable fewer. The matrix $\mathbf M$ gets promoted to an operator with eigenfunctions $V_I$ depending on the arguments of the functions, such that the expansion around a fixed point reads
\begin{equation}
 f_i(R_1,\dots,R_n;t) = f_i^\ast(R_1,\dots,R_n) + C_I V_{Ii}(R_1,\dots,R_n) \exp(-\theta_I t) \, .
\end{equation}
Here the $f_i$ are the dimensionless coupling functions, and $R_1,\dots,R_n$ are field arguments as e.g. the Ricci scalar or a scalar field.

\begin{table}[t!]
	\caption{\label{Tab.1} Summary of the mathematical structures capturing the flow of $\Gamma_k$ in different classes of approximations. Depending on the scale-dependent terms retained in $\Gamma_k$, the projected flow equations are non-linear ordinary differential equations (ODEs), partial differential equations (PDEs), or (partial) integro-differential equations (IDEs). Since fixed functionals are $k$-stationary solutions, their structure is governed by differential equations which contain one variable less than the corresponding flow equation.}
\begin{indented}
\item[]\begin{tabular}{lll} \br
approximation of $\Gamma_k$ & structure of RG flow & fixed points \\	\br% \hline \hline \hspace*{1mm}
finite number of $\cO_i$ & ODEs & algebraic \\[1.2ex]
\begin{tabular}{@{}l}
field-dependent \\
functions $f(R_1,\cdots,R_n;t)$
\end{tabular} & PDEs ($n+1$ var.) &  PDEs ($n$  var.) \\[2.4ex]
\begin{tabular}{@{}l}
momentum-dependent \\ 
form factors  $f(p_1,\cdots,p_n;t)$
\end{tabular}
& IDEs ($n+1$ var.) & IDEs ($n$ var.) \\	\br
\end{tabular}
\end{indented}
\end{table}
When studying the propagation of degrees of freedom at the quantum level, truncating the series \eref{gammaexact} at a finite number of terms may lead to spurious results. In this case, one is bound to include terms in the ansatz for $\Gamma_k$ which account for the full momentum dependence of the two-point functions. In the context of gravity, this amounts to including the two form factors introduced in equations \eref{structure1a} and \eref{structure1b} in the ansatz. When evaluating the FRGE for such an ansatz, the inclusion of these form factors generally leads to a set of non-linear integro-differential equations determining the admissible functions $W_k^{\rm C}(p^2)$, $W_k^{\rm R}(p^2)$, etc. Conceptually, this may be understood by noticing that besides the external momentum the argument of the form factor may also include the loop momentum integrated over in the trace. This entails that a consistent solution requires knowing the form factor on the entire positive real axis. Thus approximating the form factor by a Taylor series (possibly with finite radius of convergence) is not suitable and new computational methods for solving such equations are required. This will be the subject of \sref{sec:scal-tens-fluc-st-numeric}. As a concluding remark, the stability analysis is completely analogous in this case, with the difference that the operator $\mathbf M$ is now in general of integro-differential type. An overview of the different cases and their RG flow structure can be found in \tref{Tab.1}.

\subsection{The functional renormalisation group in quantum gravity}
\label{sec:FRG-qg}
Up to now, we have been abstract about organising the basis monomials $\cO_i$. 
In order to understand their structure, we first notice
that the construction of $\Gamma_k$ and its FRGE hinges on the background field formalism, at least in gauge theories and gravity. In its simplest incarnation the physical metric $g_{\mu\nu}$ is decomposed into a fixed, but arbitrary background metric $\gb_{\mu\nu}$ and fluctuations $h_{\mu\nu}$ using the linear split\footnote{Alternatively, an exponential split $g_{\mu\nu} = \gb_{\mu\alpha} [e^h]^\alpha{}_\nu$ has been considered e.g. in \cite{Kawai:1992np,Aida:1994np,Nink:2014yya,Gies:2015tca,Demmel:2015zfa,Percacci:2015wwa,Ohta:2015fcu,Alkofer:2018fxj}. In this case, the fluctuations cannot alter the signature of the physical metric, i.e., $g_{\mu\nu}$ and  $\gb_{\mu\nu}$ come with the same signature. The underlying conformal field theories arising from the exponential and linear split possess different central charges \cite{Nink:2015lmq}, suggesting that linear and exponential parameterisation lead to gravitational theories in different universality classes. For an exploration of the most general local parameterisation up to second order in the fluctuation field see \cite{Gies:2015tca}.}
\be\label{linsplit}
g_{\mu\nu} = \gb_{\mu\nu} + h_{\mu\nu} \, . 
\ee 
The trace of the fluctuation field will be denoted by $h \equiv \gb^{\mu\nu} h_{\mu\nu}$.
Conceptually, the presence of the background metric is important since it provides the metric structure which is used to discriminate between the high- and low-momentum modes. 

The background formalism also provides a convenient way to gauge-fix the freedom of performing coordinate transformations, $\delta g_{\mu\nu} = \cL_v g_{\mu\nu}$ with $\cL_v$ denoting the Lie derivative along the vector $v$. The linear split \eref{linsplit} entails that a transformation of $g_{\mu\nu}$ under diffeomorphisms may be implemented either by quantum gauge transformations,
\be
\delta^Q \gb_{\mu\nu} = 0 \, , \qquad \delta^Q h_{\mu\nu} = \cL_v g_{\mu\nu} \, , 
\ee
keeping the background metric fixed, or by background gauge transformations,
\be
\delta^B \gb_{\mu\nu} = \cL_v \gb_{\mu\nu} \, , \qquad \delta^B h_{\mu\nu} = \cL_v h_{\mu\nu} \, ,
\ee
where each quantity transforms as a tensor of the corresponding rank. By construction the gauge-fixing sector breaks the symmetry under quantum gauge transformations. At the same time, the gauge-fixing (and regulator terms) may be constructed in such a way that background gauge transformations are realised explicitly and maintained along the RG flow. The background gauge transformations may then be used to obtain a manifestly diffeomorphism invariant effective action depending on a single metric $g$ by setting $\Gamma[g] = \Gamma_{k =0}[h;\gb]|_{h=0}$.

As a result of the gauge-fixing and regulator terms, $\Gamma_k$ depends on two independent arguments which may either be chosen as $\Gamma_k[h;\gb]$ or $\Gamma_k[g,\gb]$. The former formulation has the natural interpretation of considering graviton fluctuations in a fixed but arbitrary background, while the second one emphasises the ``bi-metric'' character of the effective average action. While the two formulations are equivalent on the exact level, in typical approximations they correspond to different projections in theory space and thus yield complementary information. Both formulations allow for practical calculations: in the fluctuation language, calculations have been done resolving vertices containing up to four graviton legs \cite{Christiansen:2012rx, Codello:2013fpa, Christiansen:2014raa, Christiansen:2015rva, Christiansen:2016sjn, Denz:2016qks,Christiansen:2017bsy, Knorr:2017fus}, and first results on the all-orders fluctuation field dependence to lowest order in the momentum have been presented in \cite{Knorr:2017mhu}. In the bi-metric language, the most advanced calculations resolve both background and full metric Einstein-Hilbert structures \cite{Manrique:2010am, Becker:2014qya}. We emphasise that the object driving the flow in \eref{FRGE} is the fluctuation two-point function $\Gamma^{(2,0)}_k[h;\bar g]$. Therefore we will resort to the $h$-$\gb$-setup throughout the rest of this work.

The extra metric dependence is controlled by a split-Ward or Nielsen identity. It arises from the observation that the decomposition \eref{linsplit} is invariant under local split-symmetry transformations
\be\label{splitsymmetry}
\gb_{\mu\nu} \mapsto \gb_{\mu\nu} + \epsilon_{\mu\nu} \, , \qquad h_{\mu\nu} \mapsto h_{\mu\nu} - \epsilon_{\mu\nu}  \, . 
\ee
This transformation results in a (modified) split-Ward identity which relates functional derivatives of $\Gamma_k$ with respect to the background and fluctuation field. Schematically, it reads
\be\label{eq:NI}
\frac{\delta \Gamma_k[h;\gb]}{\delta \gb} - \frac{\delta \Gamma_k[h;\gb]}{\delta h} = \cN[h;\gb] \, ,
\ee
where $\cN[h;\gb]$ carries the information about the non-trivial behaviour of diffeomorphism transformations of the gauge-fixing sector and the regulator \cite{Litim:2002ce, Litim:2002hj, Pawlowski:2003sk, Pawlowski:2005xe, Manrique:2009uh, Manrique:2010mq, Donkin:2012ud, Bridle:2013sra, Dietz:2015owa, Safari:2015dva, Labus:2016lkh, Morris:2016spn, Morris:2016nda, Safari:2016dwj, Safari:2016gtj, Percacci:2016arh, Nieto:2017ddk, Eichhorn:2018akn}.

\begin{figure}
\centering
\usetikzlibrary{arrows,decorations.markings}
\begin{tikzpicture}[x=5cm,y=2cm,     decoration={
       markings,
       mark=at position 1 with {\arrow[scale=2,black]{latex}};
     }]
\node[draw,text width=4cm,align=center](break) at (1/2,1) {split symmetry broken by $\mathcal R_k, \Gamma^{\rm gf}_k$};
\node[draw,text width=4cm,align=center](swis) at (0,0) {solve sWIs};
\node[draw,text width=4cm,align=center](fluct) at (0,-1) {solve background flow};
\node[draw,text width=4cm,align=center](bim) at (1,0) {solve bi-metric flow};
\node[draw,text width=4cm,align=center](swisk0) at (1,-1) {solve sWIs at $k=0$};
\node[draw,text width=4cm,align=center](ssg) at (1/2,-2) {$\Gamma[\bar g] = \Gamma_0[0;\bar g]$};

\draw[postaction=decorate] (break) -- (swis);
\draw[postaction=decorate] (break) -- (bim);
\draw[postaction=decorate] (swis) -- (fluct);
\draw[postaction=decorate] (fluct) -- (ssg);
\draw[postaction=decorate] (bim) -- (swisk0);
\draw[postaction=decorate] (swisk0) -- (ssg);
\end{tikzpicture}
\caption{Setup to obtain the effective action using the two schemes outlined in the main text.}
\label{fig:splitsymmetryscheme}
\end{figure}
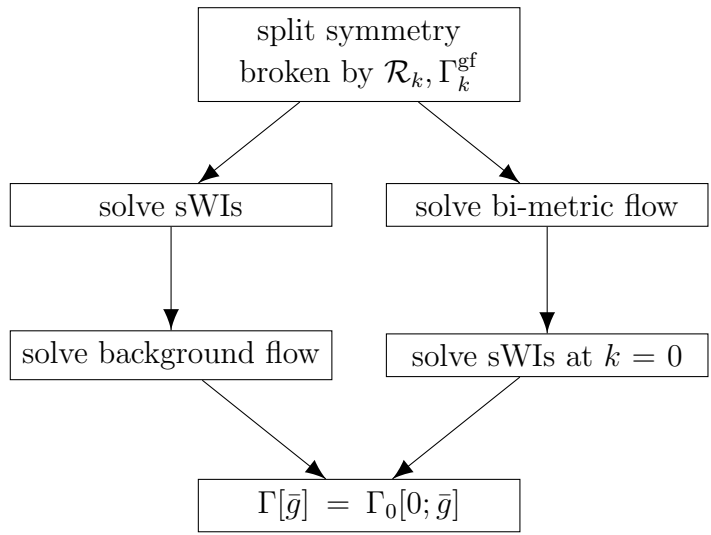

The bi-metric structure of the effective average action can be approached in two different ways, graphically depicted in \fref{fig:splitsymmetryscheme}. The first way is to solve the Nielsen identity \eref{eq:NI} by calculating the fluctuation two-point function in terms of background correlators. Structurally, the relation involves all background correlators and background derivatives of the regulator as well as the gauge fixing and ghost action,
\begin{equation}
 \Gamma_k^{(2,0)} = \Gamma_k^{(2,0)}\left[\Gamma_k^{(0,n)}[h;\bar g], \frac{\delta^n}{\delta \bar g^n} \mathcal R_k, S_{\rm{gf}}, S_{\rm{gh}} \right] \, .
\end{equation}
To lowest order, the background and fluctuation correlators agree. The corrections to this can be organised in an expansion in the number of loops. Once the approximation of the fluctuation propagator is obtained, the actual flow equation can be solved.

The second way is to employ a vertex expansion of $\Gamma_k$,
\be\label{vertexex}
\Gamma_k[h;\gb] = \sum_n \frac{1}{m!} \, \int \rmd^dx \sqrt{\gb} \left[\Gamma_k^{(m)}[\gb]\right]^{\mu_1\nu_1 \cdots \mu_m \nu_m} \, h_{\mu_1 \nu_1} \cdots h_{\mu_m \nu_m} \, .  
\ee
The correlators $\Gamma_k^{(m)}[\gb]$ depend on both background curvatures and the momenta of the fluctuation fields in terms of covariant background derivatives.
With this ansatz, one resolves both background and fluctuation correlators individually, giving rise to a higher-dimensional theory space. Clearly, these correlators are related by the Nielsen identity. The strategy most often employed is to solve this, in a sense over-complete, flow, and then to impose the Nielsen identity only in the infrared limit, $k\to0$. This is expected to reduce the dimension of theory space back to its original dimension.

The FRGE expresses the scale-dependence of $\Gamma_k^{(m)}$ in terms of $\Gamma_k^{(m+1)}$, $\Gamma_k^{(m+2)}$ and lower-order $m$-point functions. Thus solving the flow of $\Gamma_k$ at the level of the $m$-point vertex requires some closure conditions for the $(m+1)$ and $(m+2)$-point functions. The established procedure employed in most works is to retain the tensor structures of the classical action and to identify the couplings of the two highest, non-resolved correlators with couplings from the highest order resolved vertex.

%------------------------------------------------------------------
\section{Form factors for gravity and gravity-matter systems}
\label{sec:struc-funcs}
%------------------------------------------------------------------
We now proceed by discussing the momentum-dependent form factors
for gravity and gravity-matter systems. Our primary focus is on form factors obeying split symmetry and we assume that we can freely integrate by parts without generating boundary terms.
In our setting, a natural ordering principle follows from counting the number of matter fields and curvature tensors. In the context of Asymptotic Safety, matter coupled to gravity has been investigated in \cite{Percacci:2002ie, 
Percacci:2003jz, Zanusso:2009bs, Daum:2009dn, Narain:2009fy, Manrique:2010mq, Vacca:2010mj, Harst:2011zx, Eichhorn:2011pc, Folkerts:2011jz, Dona:2012am, 
Dobrich:2012nv, Eichhorn:2012va, Dona:2013qba, Henz:2013oxa, Eichhorn:2014qka, Percacci:2015wwa, Borchardt:2015rxa, Dona:2015tnf, Labus:2015ska, 
Meibohm:2015twa, Eichhorn:2016esv, Meibohm:2016mkp, Eichhorn:2016vvy, Henz:2016aoh, Christiansen:2017gtg, Biemans:2017zca, Christiansen:2017qca, 
Eichhorn:2017eht, Eichhorn:2017ylw, Wetterich:2017ixo, Eichhorn:2017lry, Eichhorn:2017sok, Christiansen:2017cxa, Eichhorn:2017egq, Eichhorn:2017muy, Eichhorn:2017als, Bonanno:2018gck, Eichhorn:2018whv, Eichhorn:2018akn, Eichhorn:2018ydy, Eichhorn:2018yfc, Pawlowski:2018ixd, Eichhorn:2018nda, Eichhorn:2019yzm, deBrito:2019epw, Wetterich:2019zdo}.
%------------------------------------------------------------------
\subsection{Form factors for split-symmetry invariant actions}
%------------------------------------------------------------------
We start by discussing the form factors determining the propagators in the matter sector in subsections~\ref{sec:struc-funcs-scal}-\ref{sec:struc-funcs-ferm} before completing the discussion with the gravitational case in subsection~\ref{sec:struc-funcs-grav}.  
%-------------------------------------------------------------------
\subsubsection{Scalars}
\label{sec:struc-funcs-scal}
%------------------------------------------------------------------
For scalar fields $\phi$, there is one form factor associated with
the kinetic term
\be\label{Sscalar}
\Gamma_k^{\rm s,kin}[\phi,g]
=
\frac{1}{2}	\int	\rmd^d x	\sqrt{g}  \, \phi	\,	f_k^{(\phi\phi)}(\Delta)	\, \phi \, . 
\ee
In analogy to computations in flat Minkowski space, we define the wave-function renormalisation of the scalar field according to \cite{Dona:2013qba}
\be\label{wavefctren}
Z_k^{\rm s} \equiv \left. \frac{\p}{\p p^2} \, \, f_k^{(\phi\phi)}(p^2) \, \right|_{p^2 = 0} \, . 
\ee
The corresponding anomalous dimension is then given by
\be\label{anomalousdim}
\eta_s(k) \equiv -	 \p_t \ln	Z_k^{\rm s} \, .
\ee
One may also extract the zero-momentum behaviour of the form factor. In this way, we define the gap parameter
\be\label{gapparameter}
\mu_k^2 \equiv (Z_k^{\rm s})^{-1} \left. \, f_k^{(\phi\phi)}(p^2) \right|_{p^2 = 0} \, .  
\ee
If $f_k^{(\phi\phi)}(p^2)$ is a linear function of the squared momentum and $\phi$ is in the symmetric phase, $\mu_k$ has the interpretation of the mass of the scalar field, since it corresponds to the pole of the (Wick-rotated) Lorentzian propagator. For a general form factor, this interpretation does not hold however and one has to analyse the pole structure of the (Wick-rotated) form factor in order to gain any insight on the masses of the propagating fields.

To linear order in the spacetime curvature, the scalar sector gives rise to two additional form factors,
\ba\label{Rphiphi}
\Gamma_k^{({\rm R}\phi\phi)}[\phi,g] = & \int \rmd^dx \sqrt{g} f_k^{({\rm R}\phi\phi)}(\Delta_1,\Delta_2,\Delta_3) \, R \phi \phi \, , \\ \label{Ricphiphi}
\Gamma_k^{({\rm Ric}\phi\phi)}[\phi,g] = & \int \rmd^dx \sqrt{g} f_k^{({\rm Ric}\phi\phi)}(\Delta_1,\Delta_2,\Delta_3) \,  R^{\mu\nu} (D_\mu D_\nu \phi) \phi \, .
\ea
Here $\Delta_i$ is the Laplacian acting on $i$-th field, $\Delta_1 (R \phi \phi) = (\Delta_1 R) \phi \phi$, cf.\ \ref{app:notation}. An investigation of correlations of this type without form factors in a bi-metric setup has been carried out in \cite{Eichhorn:2017sok}, and in a Brans-Dicke motivated context in \cite{Henz:2013oxa, Shapiro:2015ova, Henz:2016aoh,Percacci:2015wwa}. The monomials \eref{Rphiphi} and \eref{Ricphiphi} constitute a complete set of form factors at first order in the spacetime curvature. The invariant $\int \rmd^dx \sqrt{g} f_k(\Delta_1,\Delta_2,\Delta_3) \,  R^{\mu\nu} (D_\mu \phi) ( D_\nu \phi)$ can be mapped to this basis set through integration by parts and the use of the second Bianchi identity. Moreover, any pair of contracted covariant derivatives acting on different fields may be eliminated by means of the identity
\be
\fl
\int \! \rmd^dx \sqrt{g} R_1 R_2 (\Delta R_3) \! = \!\!\! \int \! \rmd^dx \sqrt{g} \left[ (\Delta R_1) R_2 + R_1 (\Delta R_2) + 2 (D_\mu R_1)(D^\mu R_2) \right] R_3 \, , 
\ee
where the $R_i$ represent arbitrary tensor fields. These manipulations typically also produce additional curvature tensors by the commutation of covariant derivatives. Since these are of higher order in $R$, they will not be considered at this stage.
%------------------------------------------------------------------
\subsubsection{Vectors}
\label{sec:struc-funcs-vec}
%------------------------------------------------------------------
The kinetic term of an abelian gauge field $A_\mu$ with field strength $F_{\mu\nu} = D_\mu A_\nu - D_\nu A_\mu$ reads
\be\label{Vkin}
\Gamma^{\rm v}_k[A_\mu,g] = \frac{1}{4} \int \rmd^dx \sqrt{g} F_{\mu\nu} \, F^{\mu\nu} \, ,
\ee
where all indices are raised and lowered with the spacetime metric $g_{\mu\nu}$. This term
can be generalised to include the form factor $f_k^{\rm v}(\Delta)$,
\be\label{vector}
\Gamma^{\rm v,kin}_k[A_\mu,g] = \frac{1}{4} \int \rmd^dx \sqrt{g} F_{\mu\nu} \, f_k^{\rm v}(\Delta) \, F^{\mu\nu} \, .
\ee
Similarly to the scalar case, $f_k^{\rm v}(x)|_{x=0}$ encodes the wave-function renormalisation for the vector fields. In the case of non-abelian gauge fields the connection $D_\mu$ is supplemented by an additional connection piece built from $A_\mu$. 

In $d=4$ spacetime dimensions there is a second interaction monomial constructed from two powers of the field strength tensor contracted with a totally antisymmetric $\epsilon$-tensor. Like for the kinetic term \eref{Vkin}, one could also generalise this term by including a form factor. In this case it turns out that the flat part is a total derivative, as is the term without form factor, whereas the commutator of the covariant derivatives will give rise to an additional field strength tensor. As a consequence the interaction monomial contains either three powers of the field strength or an additional spacetime curvature tensor and will thus not be considered here.

%------------------------------------------------------------------
\subsubsection{Fermions}
\label{sec:struc-funcs-ferm}
%------------------------------------------------------------------
Our construction of the form factors for fermionic fields builds on the spin-base formalism developed in \cite{Gies:2013noa,Lippoldt:2015cea,Gies:2015cka}. We start by introducing (spacetime-dependent) Dirac matrices $\gamma_\mu$ satisfying the Clifford algebra 
\be\label{clifford}
\{\gamma_\mu,\gamma_\nu\} = 2 g_{\mu\nu} \unit \, , \qquad \gamma_\mu \in {\rm Mat}(d_\gamma \times d_\gamma, \mathbbm{C}) \, , \quad d_\gamma = 2^{\lfloor d/2 \rfloor} \, , 
\ee
where $\unit$ denotes the unit matrix in Dirac space. Dirac fermions are then represented by a Grassmann-valued $d_\gamma$-component vector $\psi$. Fermion bi-linears are formed with the metric $\mathfrak{h}$ on Dirac space, where $\mathfrak{h} \in {\rm Mat}(d_\gamma \times d_\gamma, \mathbbm{C})$ is anti-hermitian, $\mathfrak{h}^\dagger = - \mathfrak{h}$, and has unit determinant. The conjugate of a Dirac spinor is then defined as $\bar{\psi} \equiv \psi^\dagger \mathfrak{h}$. This ensures that
$
\left( \bar{\psi} \psi \right)^\dagger = \bar{\psi} \psi  
$ 
is real. 

Since the properties of the Clifford algebra depend on the dimension of spacetime, the discussion will focus on four-dimensional (Euclidean signature) spacetimes admitting a spin-structure. In this case the $\gamma_{\mu}$ can be chosen to satisfy the reality property $(\gamma_{\mu})^\dagger = \mathfrak{h} \gamma_{\mu} \mathfrak{h}^{-1}$.  Moreover, the Clifford algebra admits an additional operator $\gamma_\ast = \frac{1}{24} \sqrt{g} \, \epsilon_{\mu\nu\rho\sigma} \gamma^\mu \gamma^\nu \gamma^\rho \gamma^\sigma$, where $\epsilon$ is the standard Levi-Civita symbol. It satisfies ${\rm tr} \gamma_\ast = 0$, $(\gamma_\ast)^\dagger = \mathfrak{h} \gamma_\ast \mathfrak{h}^{-1}$, and $(\gamma_\ast)^2 = \unit$, which allows to distinguish the left- and right-handed components of the Dirac spinor. Given a set of Dirac matrices satisfying \eref{clifford} together with the reality properties for the $\gamma_{\mu}$ uniquely determines $\mathfrak{h}$.  In order to construct a kinetic term for the fermions, we introduce a covariant derivative $\nabla_\mu$ containing the spin-base connection $\Gamma_{\mu}$,\footnote{The connection piece of $\nabla_\mu$ can be generalised to also contain a spin-torsion part $\Delta \Gamma_\mu$. This case will not be considered here.}
\be
\nabla_\mu \psi = \p_\mu \psi + \Gamma_{\mu} \psi \, , \qquad \nabla_\mu \bar{\psi} = \p_\mu \bar{\psi} - \bar{\psi} \Gamma_{\mu} \, . 
\ee
The spin-base connection $\Gamma_\mu$ is completely determined in terms of the Dirac matrices and the Levi-Civita connection,
\be
\Gamma_{\mu} = \sum_{n=1}^4 m_{\mu\rho_1 \cdots \rho_n} \gamma^{\rho_1 \cdots \rho_n} \, ,
m_{\mu\rho_1 \cdots \rho_n} \equiv \frac{(-1)^{\frac{n(n+1)}{2}} {\rm tr} \left( \gamma_{\rho_1 \cdots \rho_n} \left[ (D_\mu \gamma^\nu) , \gamma_{\nu} \right] \right)}{8 \, n! \, (4 \, (1-(-1)^n) - 2 n )} \, ,
\ee
where $D_\mu \gamma^\nu = \partial_\mu \gamma^\nu + \Gamma^\nu{}_{\mu\rho} \gamma^\rho$ and $\gamma^{\rho_1 \cdots \rho_n} = 1/n!(\gamma^{\rho_1} \cdots \gamma^{\rho_n} + \ldots)$ is the completely anti-symmetrised product of $n$ Dirac matrices. The connection ensures that $\nabla_\mu \psi$ transforms as a covector under general coordinate transformations and as a vector under ${\rm SL}(d_\gamma, \mathbbm{C})$ spin-base transformations. As an important property, $\slashed{\nabla} \equiv \gamma^\mu \nabla_\mu$ satisfies the Lichnerowicz relation
\be\label{lichnerowicz}
\Delta_{\rm D} \equiv (\mathbf i \slashed{\nabla})^2 = \left(- g^{\mu\nu} D_\mu D_\nu + \frac{1}{4} R \right) \unit \, . 
\ee

Based on these prerequisites, it is now straightforward to introduce the three independent form factors appearing at the level of fermion bi-linears,
\ba
\Gamma_k^{\rm D,1}[\bar{\psi}, \psi, g] & = & \int \rmd^4x \sqrt{g} \, \bar{\psi} \, f^{{\rm D},1}_k(\Delta_{\rm D}) \, (\mathbf i \slashed{\nabla}) \, \psi \, , \\
\Gamma_k^{\rm D,2}[\bar{\psi}, \psi, g] & = & \int \rmd^4x \sqrt{g} \, \bar{\psi} \, f^{{\rm D},2}_k(\Delta_{\rm D}) \, \gamma_\ast \,  \slashed{\nabla} \, \psi \, , \\
\Gamma_k^{\rm D,3}[\bar{\psi}, \psi, g] & = & \int \rmd^4x \sqrt{g} \, \bar{\psi} \, f^{{\rm D},3}_k(\Delta_{\rm D})  \, \psi \, , \\
\Gamma_k^{\rm D,4}[\bar{\psi}, \psi, g] & = & \int \rmd^4x \sqrt{g} \, \bar{\psi} \, f^{{\rm D},4}_k(\Delta_{\rm D}) \, \gamma_\ast \, \psi \, . 
\ea
The form factors $f^{{\rm D},1}_k(\Delta_{\rm D})$ and $f^{{\rm D},2}_k(\Delta_{\rm D})$ generalises the kinetic term. In particular, linear combinations of $f^{{\rm D},1}_k(0)$ and $f^{{\rm D},2}_k(0)$ define the wave-function renormalisations for the two chiral components of the Dirac field. The form factors $f^{{\rm D},3}_k(\Delta_{\rm D})$ and $f^{{\rm D},4}_k(\Delta_{\rm D})$ generalise the mass terms to momentum-dependent functions. Again the $k$-dependent mass of the fermion is associated with the roots of the (Lorentzian signature) Dirac equation. In a flat background the relevant equation is
\be
\bigg[   f^{{\rm D},1}_k(\Box) (\mathbf i \slashed{\p})+ f^{{\rm D},2}_k(\Box) \, \gamma_\ast \,  \slashed{\p} - \left( f^{{\rm D},3}_k(\Box) +  f^{{\rm D},4}_k(\Box) \, \gamma_\ast  \right) \bigg] \psi = 0 \, , 
\ee
 so that the construction accommodates scale-dependent mass terms for the right- and left-handed components of the Dirac fermion. Owed to the presence of the scalar curvature term in \eref{lichnerowicz} the form factors $f^{{\rm D},i}_k(\Delta_{\rm D})$ have a non-trivial overlap with scale-dependent functions $f_k(R)$ built from the Ricci scalar. A natural way to disentangle these two sets of functions is to take the flat space limit where the latter are trivial.

%------------------------------------------------------------------
\subsubsection{Gravity}
\label{sec:struc-funcs-grav}
%------------------------------------------------------------------
The first set of non-trivial form factors in the gravitational sector appears at second order in the spacetime curvature. In this case the basis for the split symmetry invariant form factors can be chosen as
\ba\label{structure2a}
\Gamma_k^{\rm C,}[g] & =  \frac{1}{16 \pi G_k}  \int \rmd^dx \sqrt{g} \, C_{\mu\nu\rho\sigma} \, W_k^{\rm C}(\Delta) \, C ^{\mu\nu\rho\sigma} \, , \\ \label{structure2b}
\Gamma_k^{\rm R}[g] & =  \frac{1}{16 \pi G_k} \int \rmd^dx \sqrt{g} \, R \, W_k^{\rm R}(\Delta) \, R \, .
\ea
The choice of these basis terms is distinguished in the sense that the form factors
$W_k^{\rm C}(\Delta)$ and $ W_k^{\rm R}(\Delta)$ lead to a non-trivial momentum dependence of the transverse-traceless and scalar propagator, respectively, when the background in \eref{linsplit} is chosen as flat Euclidean space. The third potential invariant $\int \rmd^dx \sqrt{g} \, R_{\mu\nu} \, R^{\mu\nu}$ does not give rise to an additional form factor. Any term of the structure $\int \rmd^dx \sqrt{g} \, R_{\mu\nu} \, \Delta^n \, R^{\mu\nu}$, $n \ge 1$ can be mapped to the basis elements and higher order curvature terms by means of the second Bianchi identity \eref{Bianchi12}. To demonstrate this, we first note that
\ba\nonumber
D^\alpha D_\alpha R_{\rho\sigma\mu\nu} &= - D^\alpha \left[D_\rho R_{\sigma\alpha\mu\nu} + D_\sigma R_{\alpha\rho\mu\nu}\right] \\
& = 2 D_\rho D_{[\mu} R_{\nu]\sigma} - 2 D_\sigma D_{[\mu} R_{\nu]\rho} + \cO(R^2) \, , 
\ea
where we commuted two covariant derivatives and made use of the contracted Bianchi identity \eref{Bianchi2contracted} in the second step. Contracting with a Riemann tensor, this implies
\be
R^{\rho\sigma\mu\nu} D^2 R_{\rho\sigma\mu\nu} = 4 R^{\rho\sigma\mu\nu} D_\rho D_\mu R_{\nu\sigma} + \cO(R^2)\, . 
\ee
Integrating this equation over spacetime then allows to integrate by parts, so that the covariant derivatives appearing on the right-hand side can again be arranged to act on the Riemann tensor. Again making use of the contracted Bianchi identity establishes that
\be\label{curvid}
\int \rmd^dx \sqrt{g} \left[R^{\rho\sigma\mu\nu} \Delta R_{\rho\sigma\mu\nu} - 4 R^{\mu\nu} \Delta R_{\mu\nu} + R \Delta R\right] = \cO(R^3) \, . 
\ee
This relation readily extends to higher powers of the Laplacian. This case involves additional commutators when reordering the covariant derivatives before performing the integration by parts. These additional commutators only provide further terms of order $\cO(R^3)$, so that \eref{curvid} is correct for all positive powers $\Delta^n$. Using \eref{Weyldef} in order to eliminate the Riemann tensor in favour of the Weyl tensor leads to a similar relation, albeit with different numerical coefficients,
\be
 \fl \int \rmd^dx \sqrt{g} \left[C^{\rho\sigma\mu\nu} \Delta C_{\rho\sigma\mu\nu} - 4 \frac{d-3}{d-2} R^{\mu\nu} \Delta R_{\mu\nu} + \frac{d(d-3)}{(d-1)(d-2)} R \Delta R\right] = \cO(R^3) \, . 
\ee
This establishes that the monomials \eref{structure2a} and \eref{structure2b} are indeed the only form factors that appear at second order in the curvature, see also \cite{Codello:2012kq} for related discussions.\footnote{This statement assumes that the form factors possess a well-defined (inverse) Laplace transform \eref{LaplaceTrafo}, which we tacitly assume throughout the entire work.} However, in dimensions higher than four, the Ricci-squared term without form factor has to be included, since it cannot be eliminated by the Euler characteristic.

At this stage the following remark is in order. The two monomials \eref{O01} spanning the Einstein-Hilbert action do not lend themselves to a generalisation by introducing a non-trivial form factor. Adding a function $f_k(\Delta)$ acting on the Ricci scalar in $\cO_{\rm R}$ leads to integrands which are total derivatives and thus merely contribute surface terms to the action. This case will not be considered any further at this stage.
%------------------------------------------------------------------
\subsection{Form factors in the vertex expansion of a scalar-tensor theory}
\label{sec:struc-funcs-scal-tens}
%------------------------------------------------------------------
Upon discussing the form factors of a scalar-tensor theory obeying split symmetry 
in subsection \ref{sec:struc-funcs-scal}, we now focus on the description of the same setting within the framework
of a vertex expansion. In this case, we consider fluctuations of the gravitational field $h_{\mu\nu}$ in a flat Euclidean background $\gb_{\mu\nu} = \delta_{\mu\nu}$ and vertices containing two powers of the scalar field $\phi$. Again, this setting allows to introduce momentum-dependent form factors in the interaction vertices. The flat background allows to use derivatives $\p_{i\mu}$ with $\Box_i \equiv - \p_i^2$ and momenta $p_{i\mu}$ interchangeably. Again we adopt the convention that the index $i$ indicates the field on which the derivative acts upon. We construct basis elements with the convention that derivatives acting on the last field are integrated by parts. In order to lighten our notation we drop the subscript $k$ and it is understood implicitly that all form factors also depend on the scale $k$. Where necessary, we then use a subscript to indicate the tensor structure that a given form factor is associated with.
%------------------------------------------------------------------
\subsubsection{Classification of interaction vertices in powers of \texorpdfstring{$h$}{h}}
\label{sec:struc-funcs-scal-tens-classification}
%------------------------------------------------------------------
We start by generalising the nomenclature introduced in \eref{vertexex} to the scalar-tensor case. In the presence of two fields,
\be\label{vertexex1}
\fl \Gamma_k[h,\phi;\gb] = \sum_{m,n} \frac{1}{n! \, m!} \int \rmd^dx \sqrt{\gb} \left[\Gamma_k^{(m,n)}[\gb]\right]^{\mu_1\nu_1 \cdots \mu_m \nu_m}  h_{\mu_1 \nu_1} \cdots h_{\mu_m \nu_m} \, \phi^n \, .  
\ee
The vertex entering the monomial with $m$ $h$-fields and $n$ $\phi$-fields is denoted by $\Gamma_k^{(m,n)}[\gb]$.  
Moreover, we use a vertical line followed by a string of fields to denote the projection of \eref{vertexex1} onto the corresponding string of fields. Notably, the symmetry properties in the vertices $\left[\Gamma_k^{(m,n)}[\gb]\right]^{\mu_1\nu_1 \cdots \mu_m \nu_m}$ in terms of their momentum-dependence and tensor structure appear automatically, once they are extracted from \eref{vertexex1} using the variational principle.

\mbox{} \newline
\noindent
\emph{Order $\cO(h^0)$.} The lowest order vertex describes the propagation of the scalar field in flat space and does not include the graviton fluctuation field,
\be\label{Gphiphi}
\Gamma_k|_{\phi\phi} = \frac{1}{2} \int \rmd^dx \, \phi f^{(\phi\phi)}(\Box) \phi  \, . \,  
\ee
Going to momentum space and taking the functional derivatives with respect to $\phi$ according to the prescription \eref{fctvariation} then yields the (symmetrised) two-point vertex
\be
\Gamma_k^{(0,2)}(p^2) = f^{(\phi\phi)}(p^2) \, ,
\ee
where we employed momentum conservation.

\mbox{} \newline
\noindent
\emph{Order $\cO(h^1)$.} Once terms containing powers of $h_{\mu\nu}$ are considered the vertices $\left[\Gamma_k^{(m,n)}[\gb]\right]^{\mu_1\nu_1 \cdots \mu_m \nu_m}$, $m \ge 1$, possess a non-trivial tensor structure. For $n=1$ and $m=2$ this results in four form factors $f^{(h\phi\phi)}_{\mathcal T}$ which are associated with the independent tensor structures
\be\label{eq34}
 \fl \Gamma_k|_{h\phi\phi} = \int \rmd^dx \,  \bigg[ f^{(h\phi\phi)}_{(\bar g)} \, \delta^{\mu\nu} + f^{(h\phi\phi)}_{(11)} \, \p_1^\mu \p^\nu_1 + f^{(h\phi\phi)}_{(22)} \, \p_2^\mu \p_2^\nu + f^{(h\phi\phi)}_{(12)} \, \p_1^\mu \p_2^\nu \bigg] h_{\mu\nu} \phi \phi \, .  
\ee
The fourth structure deserves a comment. While partial integration allows to write $\int \rmd^dx \, \p_1^\mu \p_2^\nu \, h_{\mu\nu} \phi \phi = - \frac{1}{2} \int \rmd^dx \, (\p_1^\mu \p_1^\nu \, h_{\mu\nu}) \phi^2$, identities of this form no longer hold if there is a form factor which contains Laplacians acing on the two $\phi$-fields with different powers. Thus, in general, the fourth tensor structure must be included.

The momentum dependence of the form factors is conveniently captured by the squares of the momenta associated with the three fields,
\be\label{momdep3pt}
f^{(h\phi\phi)}_{\mathcal T} = f^{(h\phi\phi)}_{\mathcal T}(p_1^2,p_2^2,p_3^2) \, . 
\ee
The fact that the combinations $y_{ij} \equiv p_{i\mu}\,  p_j^\mu$, $i,j = 1, \cdots, m+n$, $i \not = j$ do not appear as independent arguments follows from momentum conservation at the vertex, equation \eref{momentumconservation}, which for any three-point vertex gives
\be\label{mom3pt}
p_1^\mu + p_2^\mu + p_3^\mu = 0 \, . 
\ee
This entails 
\be
p_3^2 = (p_{1\mu} + p_{2\mu} )(p_{1}^{\mu} + p_{2}^{\mu} ) = p_1^2 + p_2^2 + 2 y_{12} \, ,
\ee
which allows to express $y_{12}$ in terms of the $p_i^2$. Identities for remaining $y_{ij}$ can be obtained along the same lines, yielding 
\be
  y_{12} = \frac{p_3^2 - p_1^2 - p_2^2}{2} \, , \quad
  y_{13} = \frac{p_2^2 - p_1^2 - p_3^2}{2} \, , \quad
  y_{23} = \frac{p_1^2 - p_2^2 - p_3^2}{2} \, .
\ee
These identities allow to eliminate the dependence of the $f_{\mathcal T}^{(h\phi\phi)}$ on the $y_{ij}$ in favour of a dependence on the squared momenta, justifying \eref{momdep3pt}.

We close the discussion with the following remarks. The classification of the form factors related to higher-order vertices follows the same pattern as the one for the three-point vertices. Firstly, one determines the independent arguments of $f_{\mathcal T}^{(h^m\phi^n)}$ using momentum conservation at the vertex. Secondly, one determines the independent tensor structures providing a suitable basis for the expansion \eref{vertexex1}. Notably, the number of tensor structures proliferates rather quickly. The vertex with $m=n=2$ is discussed in detail in \ref{app:4ptvertex}. In this case there are 59 independent tensor structures which all come with their own form factor,
\be\label{4ptsf}
f_{\mathcal T}^{(hh\phi\phi)} = f_{\mathcal T}^{(hh\phi\phi)}(p_1^2,p_2^2,p_3^2,y_{12},y_{13},y_{23}) \, .
\ee
Momentum conservation at the vertex implies that any form factor appearing in a four-point vertex can depend on six different combinations of the incoming momenta. Systematically eliminating $p_4^\mu$ by imposing momentum conservation at the vertex, $p_4^\mu = - (p_1^\mu + p_2^\mu + p_3^\mu)$, leads to the arguments appearing in \eref{4ptsf}.

The vertex expansion can readily be generalised to an arbitrary background $\gb_{\mu\nu}$. In this case, there is the additional complication that operator structures acting on the same field no longer commute. For example, the Laplacian $\Delb_1$ no longer commutes with $\Db_{1\mu} \Db_i^\mu$, $i \not = 1$. This raises the need to impose some convention on how the operators are ordered. This is particularly relevant for vertex functions of higher order, as, e.g., the four-point vertex discussed in \ref{app:4ptvertex} where the form factors depend on a subset of both $\Delb_i$ and $\Db_{i\mu} \Db_j^\mu$. In this case one may impose that all $\Db_{i\mu} \Db_j^\mu$ are to the left of all $\Delb_i$ and $\Delb_j$. Different choices for the operator ordering are equivalent up to terms of order $\mathcal O(\Rb)$.

%------------------------------------------------------------------
\subsubsection{Identities for split-symmetric vertices}
\label{sec:struc-funcs-scal-tens-split-symmetry}
%------------------------------------------------------------------
At this stage it is natural to ask about the relation between the split symmetry invariant form factors introduced in subsection \ref{sec:struc-funcs-scal} and the vertex expansion of the previous section. In order to clarify this question we expand the split symmetry invariant action \eref{Sscalar} in terms of graviton fluctuations in a flat background $\gb_{\mu\nu} = \delta_{\mu\nu}$. Following the notation introduced in subsection \ref{sec:struc-funcs-scal-tens-classification}, it is convenient to give the results in terms of projected components appearing in the expansion \eref{vertexex1}. At zeroth order in the $h$-field, the projection yields
\be\label{zerovertex}
\Gamma_k^{\rm s,kin}|_{\phi\phi}
=
\frac{1}{2}	\int	\rmd^d x \, \, \phi		f^{(\phi\phi)}(\Box)	\, \phi \, .
\ee
The terms linear in $h$ originate from expanding $\sqrt{g}$ and the form factor respectively. The former follow from 
\be
\sqrt{g} \simeq   1 + \frac{1}{2}h + \frac{1}{8} h^2 - \frac{1}{4} h^{\mu\nu} h_{\mu\nu} + \cO(h^3) \, ,
\ee
where $h \equiv \gb^{\mu\nu} h_{\mu\nu}$. The latter arise from expanding the Laplacian acting on scalar fields in powers of $h$,
\be
\Delta \, \phi \simeq \left[\Box + \mathbbm d_1 + \mathbbm d_2 + \cdots \right] \phi \, . 
\ee
The explicit computation gives
\begin{eqnarray}
\mathbbm d_1 &= h_{\mu\nu} \p^\mu \p^\nu + (\p^\mu h_{\mu\nu}) \p^\nu - \frac{1}{2} (\p_\alpha h) \p^\alpha \, , \label{eq:d1scalarm} \\
\mathbbm d_2 &= -{h_\mu}^\alpha h_{\alpha\nu} \p^\mu \p^\nu - h^{\alpha\beta}(\p_\beta h_{\alpha\mu}) \p^\mu - {h_\mu}^\beta (\p^\gamma h_{\beta\gamma}) \p^\mu \nonumber \\
&\qquad + \frac{1}{2} h^{\alpha\beta} (\p_\mu h_{\alpha\beta}) \p^\mu + \frac{1}{2} h^{\mu\nu} (\p_\mu h)\p_\nu \, . \label{eq:d2scalarm}
\end{eqnarray}
Combining these basic expansions with the computational techniques introduced in \ref{app:laplacian} allows to generalise these results to functions of the Laplacian. This results in the following expansion coefficient:
\ba\nonumber
& \Gamma_k^{\rm kin}|_{h\phi\phi}
=  \frac{1}{2}	\int	\rmd^d x \, \, \Big[ \frac{1}{2} \, f^{(\phi\phi)}(\Box_2) h \phi \phi  \\ \label{skinhpp}
& \qquad +  \int_0^\infty \!\!\!\! \rmd s \, \tilde f^{(\phi\phi)}(s) \sum_{j\geq0}^\infty \frac{(-s)^{j+1}}{(j+1)!} \sum_{l=0}^j {j \choose l} (-1)^l (\Box^{j-l} \phi) \, \mathbbm d_1 \,  \Box^l \, e^{-s\Box}\phi \Big] \, . 
\ea 
Here $\tilde{f}^{(\phi\phi)}(s)$ is the inverse Laplace transform of the form factor $f^{(\phi\phi)}(\Box)$.
Remarkably, all sums and the Laplace transform can be performed explicitly. Labelling the Laplacians acting on $h$, the first and second scalar field by $\Box_1$, $\Box_2$ and $\Box_3$, respectively, one has
\ba\nonumber
& \int_0^\infty \rmd s \, \tilde f^{(\phi\phi)}(s) \sum_{j\geq0}^\infty \frac{(-s)^{j+1}}{(j+1)!} \sum_{l=0}^j {j \choose l} (-1)^l \Box^{j-l}_2 \,   \Box^l_3 \, e^{-s\Box_3} h \phi \phi \\ \nonumber
& \quad = \int_0^\infty \rmd s \, \tilde f^{(\phi\phi)}(s) \sum_{j\geq0}^\infty \frac{(-s)^{j+1}}{(j+1)!}  \left( \Box_2  -  \Box_3 \right)^j \, e^{-s\Box_3} h \phi \phi \\ \nonumber
& \quad = \left( \Box_2 \, -  \Box_3 \right)^{-1} \int_0^\infty \rmd s \, \tilde f^{(\phi\phi)}(s) \left( e^{-s(\Box_2  - \Box_3)} - 1 \right) \, e^{-s\Box_3} h \phi \phi \\
& \quad = \left( \Box_2 \, -  \Box_3 \right)^{-1} \, \left( f^{(\phi\phi)}(\Box_2) - f^{(\phi\phi)}(\Box_3) \right)  h \phi \phi \, . 
\ea
Substituting this result into \eref{skinhpp} then yields the final form of the coefficient $\Gamma_k^{\rm s,kin}|_{h\phi\phi}$:
\ba\nonumber
& \Gamma_k^{\rm s,kin}|_{h\phi\phi}
=  \frac{1}{2}	\int	\rmd^d x \, \, \Big[ \frac{1}{2} \, \delta^{\mu\nu} f^{(\phi\phi)}(\Box_2) \\  \label{skinhpp2}
&  + \frac{f^{(\phi\phi)}(\Box_2) - f^{(\phi\phi)}(\Box_3)}{ \Box_2 \, -  \Box_3} \, 
\left(\p^\mu_3 \p^\nu_3 + \p^\mu_1  \p^\nu_3 - \frac{1}{2} \delta^{\mu\nu} \p_{1\alpha}  \p^\alpha_3 \right)
\Big] h_{\mu\nu} \,\phi \phi \, . 
\ea
A series expansion of the second line shows that the coefficient is finite also on the locus $\Box_2 \, -  \Box_3 = 0$.

The expansion of the monomials \eref{Rphiphi} and \eref{Ricphiphi} in metric fluctuations in a flat background starts at order $h$. Thus they do not contribute to $\Gamma_k|_{\phi\phi}$. The terms $\Gamma^{({\rm R}\phi\phi)}_k|_{h\phi\phi}$ and $\Gamma_k^{({\rm Ric}\phi\phi)}|_{h\phi\phi}$ are found by replacing the curvature tensors by their leading coefficients in $h$. They read
\be\label{expRphiphi}
\Gamma^{{\rm R}\phi\phi}_k|_{h\phi\phi} = \int \rmd^dx f^{({\rm R}\phi\phi)}(\Box_1,\Box_2,\Box_3) \left[\Box_1 \delta^{\mu\nu} + \p_1^\mu \p_1^\nu \right]
h_{\mu\nu}\phi\phi \, , 
\ee
and
\ba\nonumber
\Gamma^{{\rm Ric}\phi\phi}_k|_{h\phi\phi} = & \frac{1}{2} \int \rmd^dx \, f^{({\rm Ric}\phi\phi)}(\Box_1,\Box_2,\Box_3) \,  \Big[
 (\Box_1 + \Box_2-\Box_3) \, \p^\mu_1 \, \p^\nu_2 \\ \label{expRicphiphi}
 & \qquad - \frac{1}{4} (\Box_1 + \Box_2-\Box_3)^2 \,  \delta^{\mu\nu} + \Box_1 \, \p^\mu_2 \, \p^\nu_2 \Big]
h_{\mu\nu}\phi\phi \, .
\ea

The structure of the four-point vertex can be analysed along the same lines. Since the intermediate steps and results are bulky, they have been relegated to \ref{app:4ptvertex}. At this stage, it is sufficient to remark that split-symmetric actions containing more than one curvature tensor do not contribute to the $(h\phi\phi)$-vertex when expanded around a flat background. Thus \eref{skinhpp2}, \eref{expRphiphi} and \eref{expRicphiphi} capture all contributions originating from a split-symmetric action.

We are now in the position to discuss the relation between the split symmetry invariant interaction monomials and the vertex expansion discussed in subsections \ref{sec:struc-funcs-scal} and \ref{sec:struc-funcs-scal-tens-classification}, respectively. In general, any split-symmetric action will give rise to an infinite tower of interaction vertices $\Gamma_k^{(m,n)}$ when one expands in graviton fluctuations $h$ around a background $\gb$. This tower then entails that there must be a relation between the form factors appearing in the split-symmetric action and the vertex expansion.\footnote{A similar relation also holds once terms breaking split symmetry are added, as the regulator and the gauge fixing term. These are controlled by the Nielsen identity \eref{eq:NI}. The modifications induced by these terms will not be discussed here.} We illustrate this general property for the form factor appearing in the scalar kinetic term \eref{Sscalar}. Combining \eref{zerovertex} and \eref{skinhpp2} gives the expansion of the split-symmetric scalar kinetic term $\Gamma_k^{\rm s,kin}[\phi,g]$  in a flat background spacetime up to terms of second order in $h$:
\ba\label{skinexp}
& \Gamma_k^{\rm s,kin} = \frac{1}{2}	\int	\rmd^d x \, \, \Big\{   \phi f^{(\phi\phi)}(\Box)	\, \phi  + \Big[ \frac{1}{2} \, \delta^{\mu\nu} 		f^{(\phi\phi)}(\Box_2) \\ \nonumber
&  + \frac{f^{(\phi\phi)}(\Box_2) - f^{(\phi\phi)}(\Box_3)}{ \Box_2 \, -  \Box_3} \, 
\big(\p^\mu_2 \p^\nu_2 + \p^\mu_1  \p_2^\nu - \frac{1}{2} \delta^{\mu\nu} \p_{1\alpha}  \p^\alpha_2 \big) \Big]
 h_{\mu\nu} \,\phi \phi + \cO(h^2) \Big\} \, .  
\ea  
Here we used the symmetry in the two $\phi$-fields to exchange the indices $2$ and $3$. By comparing \eref{skinexp} to \eref{Gphiphi} and \eref{eq34}, we see that split symmetry entails a specific relation between the form factors appearing in the vertex expansion. The $(\phi\phi)$-vertex receives contributions from the scalar kinetic term only, and anticipating this we deliberately chose the same name for the two functions.

At the level of the $(h\phi\phi)$-vertices, the split-symmetric expansions \eref{skinhpp2}, \eref{expRphiphi} and \eref{expRicphiphi} induce the following form factors associated with the tensor structures \eref{eq34}:
\ba\nonumber
f_{(\bar g)}^{(h\phi\phi)} &= \frac{1}{8} \left[ f^{(\phi\phi)}(p_2^2) + f^{(\phi\phi)}(p_3^2) - p_1^2 \frac{f^{(\phi\phi)}(p_2^2)-f^{(\phi\phi)}(p_3^2)}{p_2^2-p_3^2} \right] + p_1^2 f^{({\rm R}\phi\phi)}
\\ & \label{split1}
- \frac{1}{8} (p_1^2 + p_2^2- p_3^2)^2 f^{({\rm Ric}\phi\phi)}
\, , \\ \label{split2}
f_{(11)}^{(h\phi\phi)} &= \, f^{({\rm R}\phi\phi)} \, , \\ \label{split3}
f_{(22)}^{(h\phi\phi)} &= \frac{1}{2} \frac{f^{(\phi\phi)}(p_2^2) - f^{(\phi\phi)}(p_3^2)}{p_2^2 - p_3^2} + \frac{1}{2} p_1^2 \, f^{({\rm Ric}\phi\phi)} \, , \\ \label{split4}
f_{(12)}^{(h\phi\phi)} &= \frac{1}{2} \frac{f^{(\phi\phi)}(p_2^2) - f^{(\phi\phi)}(p_3^2)}{p_2^2 - p_3^2} + \frac{1}{2} (p_1^2 + p_2^2- p_3^2) \, f^{({\rm Ric}\phi\phi)} \, .
\ea
To ease the notation we have suppressed the arguments of all functions which depend on all three squared momenta. This establishes that split symmetry enforces relations between the four independent form factors appearing in the vertex expansion. In other words, extracting the parts of the $h\phi\phi$-vertices which can be completed into split-symmetric actions requires contributions from all four tensor structures with the corresponding momentum-dependent form factors being fixed in terms of the two free functions $f^{({\rm R}\phi\phi)}$ and $f^{({\rm Ric}\phi\phi)}$ by the relations \eref{split1}-\eref{split4}. This also establishes that there are two combinations of tensor structures which cannot be completed into split-symmetric monomials. At the practical level this suggests that the amount of split symmetry breaking induced by the regulator and gauge fixing terms can be quantified by these equations. This provides a much more straightforward way to check how strongly the full diffeomorphism symmetry is broken than the evaluation of the non-trivial Nielsen identity.

Similar relations for the form factors associated with vertices of higher order $f_{\mathcal T}^{(h^m\phi^2)}$, $m \ge 2$, can be obtained along the same lines. Their explicit construction requires classifying all split-symmetric invariants containing up to $m$ powers of the curvature tensor. Subsequently, this set of actions is expanded in the fluctuation field up to $m$-th order. The result is then compared to the classification of tensor structures involving $m$ $h$-fields and two scalars. Given that $f_{\mathcal T}^{(hh\phi\phi)}$ already gives rise to 59 independent tensor structures, it is clear that already the next order of relations will be very involved. We relegate partial results to \ref{app:4ptvertex}.

We close this section with a conceptual remark. In general the effective average action $\Gamma_k$ doesn't reduce to a functional of only one metric in the limit $k \to 0$. The reason is that the Nielsen identity is still non-trivial even in this limit, in particular because the gauge-fixing term is still present. Our discussion in terms of form factors then might serve as an approximate solution to the Nielsen identity which doesn't include the breaking induced by gauge fixing, and fix an infinite number of couplings. In the specific example of the $h\phi\phi$-vertices, split symmetry restoration (up to the non-trivial part of the Nielsen identity) at $k=0$ corresponds to imposing the boundary conditions \eref{split1}-\eref{split4}, thereby fixing two linear combinations of the form factors introduced in \eref{eq34}.

%------------------------------------------------------------------
\section{Momentum-dependent propagators in the scalar-tensor model}
\label{sec:scal-tens}
%------------------------------------------------------------------
Upon surveying the conceptual properties of momentum-dependent form factors we now focus on the computational techniques which actually allow to determine them as solutions of the Wetterich equation \eref{FRGE}. For clarity, we focus on the simplest case and consider the form factor associated with the scalar kinetic term \eref{Sscalar} with the gravitational sector approximated by the gauge-fixed Einstein-Hilbert action. In subsection \ref{sec:scal-tens-setup} we present the setup, whereas in subsection \ref{sec:scal-tens-IDE} we give the explicit flow equations. Our results are summarised in subsection \ref{sec:scal-tens-fluc-st}. The technical implementation of the derivation is detailed in the supplementary notebook {\tt flowequation.nb}.
%------------------------------------------------------------------
\subsection{Setup}
\label{sec:scal-tens-setup}
%------------------------------------------------------------------
We study the flow of the form factor of a scalar field coupled to gravity. As explained in the previous section, we approximate the flow of the three- and four-point vertex based on a diffeomorphism-invariant ansatz for the effective average action:
\begin{equation}\label{gammaans}
			\Gamma_k[g,\phi,\bar{c},c;\bar{g}]
	\approx
			\Gamma_k^{\mathrm{grav}}[g]
		+	\Gamma_k^{\mathrm{scalar}}[\phi,g]
		+	\Gamma_k^{\mathrm{gf}}[g;\bar{g}]
		+	S^{\mathrm{gh}}[g,\bar{c},c;\bar{g}]
	\,.
\end{equation}
The gravitational part is taken to be the Einstein-Hilbert action
\begin{equation}
			\Gamma_k^{\mathrm{grav}}[g]
	=
			\frac{1}{16\pi G_k}	\int\rmd^d x	\sqrt{g}	\left[
				2	\Lambda_k
			-	R
			\right]
	\, .
\end{equation}
This action includes a scale-dependent Newton's coupling $G_k$ and a cosmological constant $\Lambda_k$. The gravitational action is accompanied by a gauge-fixing action $\Gamma_k^{\mathrm{gf}}$ and a ghost term $S^{\mathrm{gh}}$. We implement the gauge fixing by the harmonic gauge
\begin{equation}
			\Gamma_k^{\mathrm{gf}}
	=
			\frac{1}{32\pi G_k}	\int	\rmd^d x	\sqrt{\bar{g}}	\mathcal{F}_\mu	\bar{g}^{\mu\nu}	\mathcal{F}_\nu
	\,,
	\qquad
			\mathcal{F}_\nu
	=
			\bar{D}^\mu	h_{\mu\nu}
		-	\frac{1}{2}	\bar{D}_\nu	h
	\,.
\end{equation}
The gauge fixing gives rise to the standard ghost-term
\begin{equation}
			\fl S^{\mathrm{gh}}
	=
		-	\sqrt{2}	\int	\rmd^d x \sqrt{\bar{g}}	\bar{c}_\mu	\left[
				\bar{D}^\rho	\bar{g}^{\mu\sigma}	g_{\sigma\nu}	D_\rho
			+	\bar{D}^\rho	\bar{g}^{\mu\sigma}	g_{\rho\nu}	D_\sigma
			-	\bar{D}^\mu	\bar{g}^{\rho\sigma}	g_{\rho\nu}	D_\sigma
			\right]	c^\nu
	\,.
\end{equation}
Finally, the form factor of the real scalar field is given by
\begin{equation}\label{eq.73}
			\Gamma_k^{\mathrm{scalar}}[\phi,g]
	=
			\frac{1}{2}	Z_k^s	\int	\rmd^d x	\sqrt{g} \,	\phi	\bar{f}_k(\Delta)	\phi
	\,,
\end{equation}
where $\Delta \equiv - g^{\mu\nu} D_\mu D_\nu$ is the Laplacian constructed from the full metric. The couplings are parameterised with a scale-dependent wave-function renormalisation $Z_k$; the strength of the gravitational interaction is encoded in the function $\bar{f}_k$, which is subject to the constraint
\be
			\bar{f}'(0)
	=
			1
	\,.
\ee

\begin{figure}
	\centering
	\includegraphics{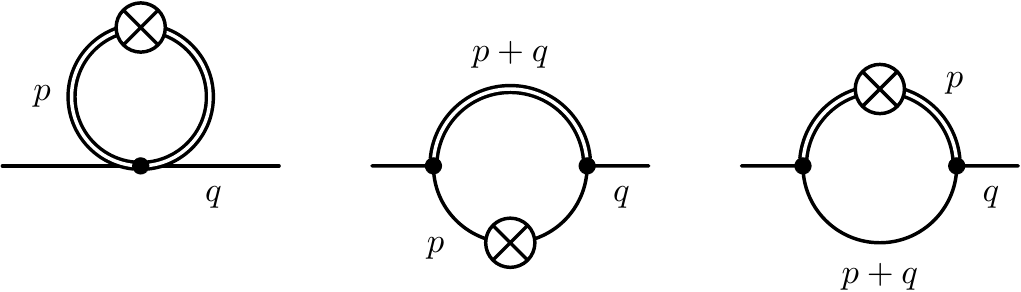}
	\caption{Feynman diagrams contributing to the flow of the scalar kinetic form factor. The solid and double lines denote the scalar and the graviton propagator, respectively. The crossed circle denotes the insertion of the cutoff operator $\partial_t \mathcal{R}_k$.
}
	\label{fig:feynman}
\end{figure}

This gravity-matter system is expected to possess a NGFP suitable for asymptotic safety \cite{Percacci:2002ie,Dona:2013qba,Meibohm:2015twa,Becker:2017tcx}. This fixed point is already visible in the simplest projection where $\Gamma_k^{\rm scalar}[\phi,g]$ is approximated by the classical action of a minimally coupled scalar field, approximating 
\be\label{approx1}
Z_k = 1 \, , \qquad \bar{f}_k(\Delta) = \Delta \, .
\ee
For $d=4$ it is situated at positive values of Newton's coupling and the cosmological constant and exhibits a complex pair of critical exponents with a positive real part. Hence the fixed point acts as a UV-attractor for the RG flow in the $G$-$\Lambda$--plane. It is connected to the one found for pure gravity through an analytic continuation in the number of scalar fields.

%------------------------------------------------------------------
\subsection{Flow equations}
\label{sec:scal-tens-IDE}
%------------------------------------------------------------------
We now present the beta functions of the scalar-tensor system resulting from the ansatz \eref{gammaans}. These are conveniently expressed in terms of their dimensionless counterparts. For the gravitational couplings, we define
\begin{equation}
			g_k
	\equiv
			k^2	G_k
	\,,\qquad
			\lambda_k
	\equiv
			k^{-2}	\Lambda_k
	\,,
\end{equation}
whereas we have the functional rescaling
\begin{equation}
			\bar{f}(z)
	\equiv
			k^{-2}	f_k\left(k^2 z\right)
	\,.
\end{equation}
On top of that, we define the gravitational and scalar anomalous dimensions, respectively:
\begin{equation}
			\eta_N
	\equiv
			\left(	G_k	\right)^{-1} \p_t	G_k
	\,,\quad
			\eta_s
	\equiv
		-	\left(	Z_k	\right)^{-1} \p_t	Z_k
	\,.
\end{equation}
For convenience, we introduce the dimensionless regulator shape function $r$ by
\begin{equation}
 R_k(z) = k^2 \, r\left(\frac{z}{k^2}\right) \, ,
\end{equation}
where $R_k$ is the scalar part of the regulator $\mathcal R_k$.

\subsubsection{Gravitational beta functions}
The beta functions in the gravitational sector are explicitly given by
\begin{equation}\label{betag}
			\partial_t	g
	=
			\left(	d	-	2	+	\eta_N	\right)	g
	\, ,
\end{equation}
\begin{equation}\label{betalambda}
			\partial_t	\lambda
	=
			g	\left(
				L_1(\lambda)
			+	L_3[f]
			\right)
		+	g	\eta_N	L_2(\lambda)
		-	(2-\eta_N)	\lambda
	\,.
\end{equation}
The anomalous dimension $\eta_N$ can be cast into the form
\begin{equation}
			\eta_N
	=
			\frac{g	\left(	B_1(\lambda)	+	B_3[f]	\right)}{1	-	g	B_2(\lambda)}
	\,.
\end{equation}
The functions $B_1(\lambda)$, $B_2(\lambda)$, $L_1(\lambda)$ and $L_2(\lambda)$ only depend on the gravitational sector, and were first derived in \cite{Reuter:1996cp}. Explicitly, they are given by
\begin{eqnarray}
	\eqalign{
			B_1(\lambda)
		=&
			\frac{1}{3}	(4\pi)^{1-d/2}	\bigg(
				d(d+1)	\Phi^1_{d/2-1}[\ell_{-2\lambda}]
			-	4d	\Phi^1_{d/2-1}[\ell_0]
	\\&
			-	6d(d-1)	\Phi^2_{d/2}[\ell_{-2\lambda}]
			-	24	\Phi^2_{d/2}[\ell_0]
			\bigg) \, ,
	}
	\\
	\eqalign{
			B_2(\lambda)
		=&
		-	\frac{1}{6}	(4\pi)^{1-d/2}	\bigg(
				d(d+1)	\tilde{\Phi}^1_{d/2-1}[\ell_{-2\lambda}]
			-	6 d	\tilde{\Phi}^2_{d/2}[\ell_{-2\lambda}]
			\bigg) \, ,
	}
	\\
	\eqalign{
			L_1(\lambda)
		=&
			(4\pi)^{1-d/2}	\bigg(
				d(d+1)	\Phi^1_{d/2}[\ell_{-2\lambda}]
			-	4d	\Phi^1_{d/2}[\ell_0]
			\bigg) \, ,
	}
	\\
	\eqalign{
			L_2(\lambda)
		=&
		-	\frac{1}{2}	(4\pi)^{1-d/2}	d(d+1)	\tilde{\Phi}^1_{d/2}[\ell_{-2\lambda}] \, .
	}
\end{eqnarray}
The functionals $B_3$ and $L_3$ are novel, and read
\begin{eqnarray}
			B_3[f]
	=&
			\frac{2}{3}	(4\pi)^{1-d/2}	\left(
				\Phi^{1}_{d/2-1}[f]
			-	\frac{1}{2}	\eta_s \,	\tilde{\Phi}^{1}_{d/2-1}[f]
			\right)
	\,,
	\\
			L_3[f]
	=&
			(4\pi)^{1-d/2}	\left(
				\Phi^{1}_{d/2}[f]
			-	\frac{1}{2}	\eta_s \,	\tilde{\Phi}^{1}_{d/2}[f]
			\right)
	\,.
\end{eqnarray}
In these expressions, we have conveniently used the generalised threshold functionals
\begin{eqnarray}
			\Phi^{p}_n[f]
	=&
			\frac{1}{\Gamma(n)}	\int_0^\infty	\rmd z	z^{n-1}	\frac{r(z)-z r'(z)}{(f(z)+r(z))^p}
	\,,
	\\
			\tilde{\Phi}^{p}_n[f]
	=&
			\frac{1}{\Gamma(n)}	\int_0^\infty	\rmd z	z^{n-1}	\frac{r(z)}{(f(z)+r(z))^p}
	\,.
\end{eqnarray}
These functionals reduce to the threshold functionals as defined in \cite{Reuter:1996cp} for a function $\ell_w$ of the form $\ell_w(z) = z +w$:
\begin{equation}
			\Phi^p_n(w)	=	\Phi^{p}_n[\ell_w]
	\,,\quad
			\tilde{\Phi}^p_n(w)	=	\tilde{\Phi}^{p}_n[\ell_w]
	\,.
\end{equation}

\subsubsection{Propagator beta function}
The flow equation for $f$ is given by
\begin{equation}\label{structuremaster}
			\left(	1	-	\frac{1}{2}	\eta_s	\right)	f(q^2)	
		+	\frac{1}{2}	\partial_t	f(q^2)
		-	q^2	f'(q^2)
	=
			\mathcal{K}_1
		+	\mathcal{K}_2
		+	\mathcal{K}_3
	\,,
\end{equation}
where the $\mathcal{K}_i$ correspond to the three Feynman diagrams in \fref{fig:feynman}. The diagram consists structurally of a momentum and angular integral over a number of graviton and scalar propagators, and the derivative of the regulator, connected by their vertex functions. The structural part of the diagrams reads
\begin{eqnarray}\eqalign{
			\mathcal{K}_1
	=
		-	32	g	\int \rmd \mu[\eta_N] \,	\mathcal{V}_1(p,q,x)	\left(	G_0^{\mathrm{grav}}(p^2)	\right)^2
	\, ,}
	\\\eqalign{
			\mathcal{K}_2
	=
			32	g	\int \rmd \mu[\eta_s] \,	\mathcal{V}_2(p,q,x)	G_0^{\mathrm{grav}}(\mathfrak{s})	\left(	G_0^{\mathrm{scalar}}(p^2)\right)^2	
	\, ,}
	\\\eqalign{
			\mathcal{K}_3
	=
			32	g	\int \rmd \mu[\eta_N] \,	\mathcal{V}_3(p,q,x)	\left(	G_0^{\mathrm{grav}}(p^2)	\right)^2	G_0^{\mathrm{scalar}}(\mathfrak{s})
	\,,
}\end{eqnarray}
where we have introduced the regularised integrals
\begin{equation}
\fl
	\int \rmd \mu[\eta]
	\equiv
			\frac{4\pi^{d/2}}{\Gamma(d/2)}	\int_0^\infty	\rmd p \int_{-1}^1	\rmd x \, 	(1-x^2)^{\frac{d-3}{2}}	\left[
		    r\left(p^2\right)	-	\frac{1}{2}	\eta	r\left(p^2\right)
			- p^2	r'\left(p^2\right)
			\right] \, ,
\end{equation}
for $\eta \in \{\eta_s, \eta_N\}$. The regularized dimensionless graviton and matter propagators are
\begin{eqnarray}
 G_0^{\mathrm{grav}}(z) = \left(z+r(z)-2\lambda\right)^{-1} \, , \\ \label{Gscalar} G_0^{\mathrm{scalar}}(z)=\left(f(z)+ r(z) \right)^{-1} \, .
\end{eqnarray}
  We have also introduced the Mandelstam variable $\mathfrak{s}=p^2+q^2+2pqx$.
The vertex functions $\mathcal{V}_i$ are given by
\begin{eqnarray}
	\eqalign{
	\fl		\mathcal{V}_1
		=
			-	\frac{1}{8}	d	(d+1)	f(q^2)
			+	\frac{1}{2}	f'\left(q^2\right)	q^2	
					\left(
						d
					-	\frac{\mathfrak{s}-p^2+\frac{d-4}{d-2}q^2}{\mathfrak{s}-q^2}
					\right)	
		\\
			+	\frac{1}{2}	\frac{f\left(\mathfrak{s}\right)-f\left(q^2\right)}{\mathfrak{s}-q^2}	q^2	\left[
					1
				+	\frac{\mathfrak{s}-p^2+\frac{d-4}{d-2}q^2}{\mathfrak{s}-q^2}
				\right]
		\, ,
	}
	\\
	\eqalign{
	\fl		\mathcal{V}_2
		=
			-	\frac{d}{2}	\frac{1}{d-2}	\left(	f(p^2)	\right)^2
			+	\frac{1}{2}	\left(	\mathfrak{s}	+	\frac{d+2}{d-2}p^2	-	q^2	\right)	f(p^2)	\frac{f(p^2)-f(q^2)}{p^2-q^2}
		\\
			-	\left(	\frac{\mathfrak{s}}{2} +	\frac{1}{d-2}	p^2	-q^2	\right)	\left(	\frac{f(p^2)-f(q^2)}{p^2-q^2}	\right)^2	p^2	
		\, ,
	}
	\\
	\eqalign{
	\fl		\mathcal{V}_3
		=
			-	\frac{d}{2}	\frac{1}{d-2}	\left(	f(\mathfrak{s})	\right)^2
			+	\frac{1}{2}	\left(	p^2	+	\frac{d+2}{d-2}\mathfrak{s}	-	q^2	\right)	f(\mathfrak{s})	\frac{f(\mathfrak{s})	-	f(q^2)}{\mathfrak{s}-q^2}
		\\
			-	\left(	\frac{p^2}{2}	+	\frac{1}{d-2}\mathfrak{s}	-	q^2	\right)	\left(\frac{f(\mathfrak{s})-f(q^2)}{\mathfrak{s}-q^2}\right)^2	\mathfrak{s}
		\,.
	}
\end{eqnarray}
Note that the vertices $\mathcal{V}_2$ and $\mathcal{V}_3$ are related by the exchange $\mathfrak{s} \leftrightarrow p^2$, consistent with the fact that the diagrams differ only in the insertion of the regulator $\partial_t \mathcal{R}_k$ and a relabeling of momenta.

We remark that the left-hand side of the beta function is a function of $q^2$, whereas on the right-hand side also odd powers of $q$ occur. We can properly symmetrise the expression for the right-hand side by the observation that odd powers of $x$ integrate to zero.

From this equation, we derive a separate equation for $\eta_s$. This is done by taking two derivatives with respect to $q$ and consecutively setting $q=0$. Then using the requirement that $f'(0)=1$ gives the following implicit equation for $\eta_s$:
\begin{equation}
	\eqalign{
			\eta_s
		=
		-	\left.\frac{\rmd^2}{\rmd q^2}	\left(
				\mathcal{K}_1	+	\mathcal{K}_2	+	\mathcal{K}_3
			\right)	\right|_{q=0, f'(0)=1} \, .
	}
\end{equation}
The explicit expression for $\eta_s$ is rather lengthy and can be found in the supplementary notebook {\tt fpsolver.nb}.

%------------------------------------------------------------------
\subsection{Fixed point solution}
\label{sec:scal-tens-fluc-st}
%------------------------------------------------------------------
In the following we will discuss the solution of the set of flow equations presented above. We will focus on the fixed point where all couplings are independent of the renormalisation group scale $k$. We begin by analysing the analytic structure. Afterwards, we present our numerical techniques to solve the equations. With this at hand, we finally present the numerical solution to the fixed point equations. The details of the numerical analysis can be found in the supplementary notebook {\tt fpsolver.nb}.
%------------------------------------------------------------------
\subsubsection{Analytic properties}
\label{sec:scal-tens-fluc-st-analytic}
We study the asymptotic properties of the fixed point solution by the assertion that for large momentum $q^2$, the function $f$ behaves like $f(q^2) \sim f_\infty q^{2\alpha}$, where $f_\infty, \alpha>0$. Inserting this into the fixed point equation gives a consistent equation for $\alpha<2$. In this range, the diagram $\mathcal{K}_2$ is always sub-leading. 

Explicitly, the equation for $\alpha$ reads
\begin{equation}\label{eq:alpha}
	\eqalign{
		\fl&	-	\frac{1}{2}	\eta_s	-	(\alpha-1)	=
	\\\fl&\qquad\quad	\frac{4	g (d-3)	(d-2\alpha)	(d-2(\alpha-1))}{(4\pi)^{d/2} \, (d-2)}			\left[	\Phi^2_{d/2}[\ell_{-2\lambda}]	-	\frac{1}{2}	\eta_N	\tilde{\Phi}^2_{d/2}[\ell_{-2\lambda}]	\right]
	\,.
	}
\end{equation}
We can solve this quadratic equation for the asymptotic exponent $\alpha$ in terms of $\eta_s$, $g$ and $\lambda$. Expanding in $g$, we infer that $\alpha$ is of the form
\begin{equation}
			\alpha
	=
			1	-	\frac{1}{2}\eta_s	+	\mathcal{O}(g)
	\,,
\end{equation}
which is in agreement with the classical expectation that the fall-off behaviour of the propagator is determined by the anomalous dimension and receives quantum corrections due to gravity.
Interestingly, we note that the $\mathcal O(g)$ corrections in \eref{eq:alpha} in $d=3$ dimensions vanish exactly.

Another analytic property that we can derive is the small-momentum behaviour of $f$. Upon evaluation at $q=0$, the fixed point equation reduces to the form
\begin{equation}\label{eq:zeromomentum}
			\left(1-\frac{1}{2}	\eta_s	\right)	f(0)
	=
			\mathcal{L}_0(g,\lambda,f)	f(0)
	\, ,
\end{equation}
We conclude that this equation is satisfied if $f(0)=0$, or if $\mathcal{L}_0 = 1-\frac{1}{2}\eta_s$. Given a numerical solution, we can check whether one or both of these conditions holds.

%------------------------------------------------------------------
%------------------------------------------------------------------
\subsubsection{Complete solutions through pseudo-spectral methods}
\label{sec:scal-tens-fluc-st-numeric}
%------------------------------------------------------------------
In this section, we present the techniques that we used in our numerical study of the fixed point equation. The fixed point equation is a non-linear integro-differential equation, and cannot be solved exactly. Therefore, we resort to pseudo-spectral numerical methods \cite{Boyd:ratCheb, Boyd:ChebyFourier, Borchardt:2015rxa, Borchardt:2016pif} to obtain an approximate solution.

\begin{figure}
	\centering
	\includegraphics[width=.6\textwidth]{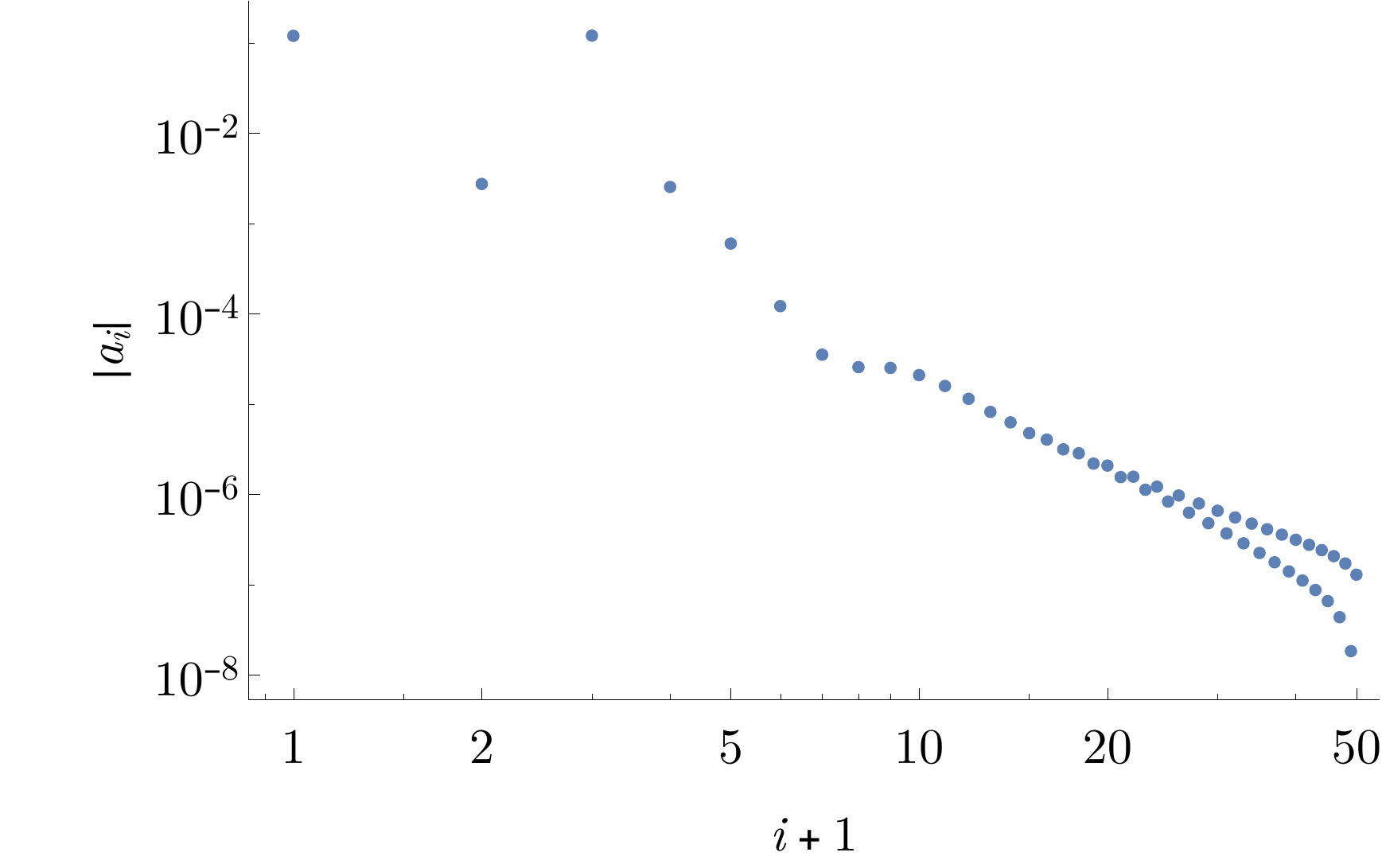}
	\caption{Absolute values of the Chebyshev coefficients $a_i$. The coefficients show algebraic convergence up to an accuracy of $10^{-6}$.}
	\label{fig:coeffs}
\end{figure}

The solving algorithm is implemented as follows. First, in order to obtain a bounded function on a compact domain, we rescale the function $f$ as
\[
			f(z)	=	(1+z)^{\alpha_{\rm num}}	\tilde{f}\left(	\frac{z-L}{z+L}	\right)
	\,,
\]
where we take the compactification scale $L$ to be $1$. From the previous paragraph, we see that setting $\alpha_{\rm num} = 2$ is sufficiently large that the function $\tilde{f}$ is bounded. The function $\tilde{f}$ is now expanded in Chebyshev polynomials,
\[
			\tilde{f}(s)
	=
			\sum_{i=0}^{N-1}	a_i	T_i(s)
	\,,
\]
where $T_i$ denotes the $i$-th Chebyshev polynomial of the first kind. Evaluation of $\tilde{f}$ and its derivatives is efficiently achieved by a Clenshaw algorithm. By evaluating the fixed point equation on the Gauss-Chebyshev grid of degree $N$, we obtain $N$ independent algebraic equations for the $a_i$.

The constraints for $f'(0)=1$ and $\eta_s$ are implemented by replacing the equations of the first two collocation points. Together with the gravitational beta functions and the constraint $\eta_N=2-d$, this yields a set of $N+4$ equations for the $N+4$ parameters $\{ g, \lambda, \eta_N, \eta_s\} \cup \{a_i\}_{i=0}^{N-1}$.
Solving this set of equations is achieved by a Newton-Raphson algorithm.

\begin{figure}
	\centering
	\includegraphics[width=.6\textwidth]{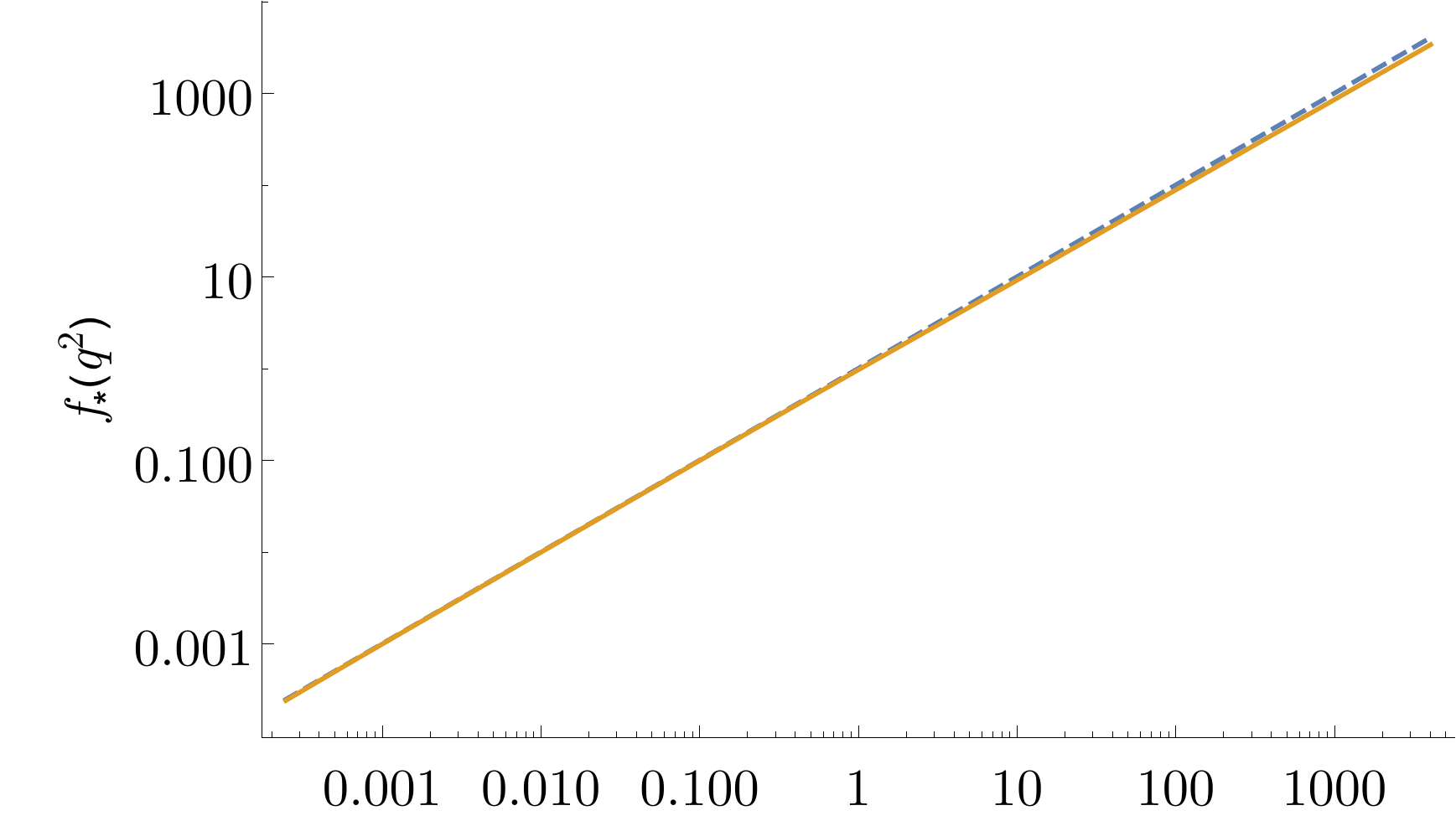}
	\includegraphics[width=.6\textwidth]{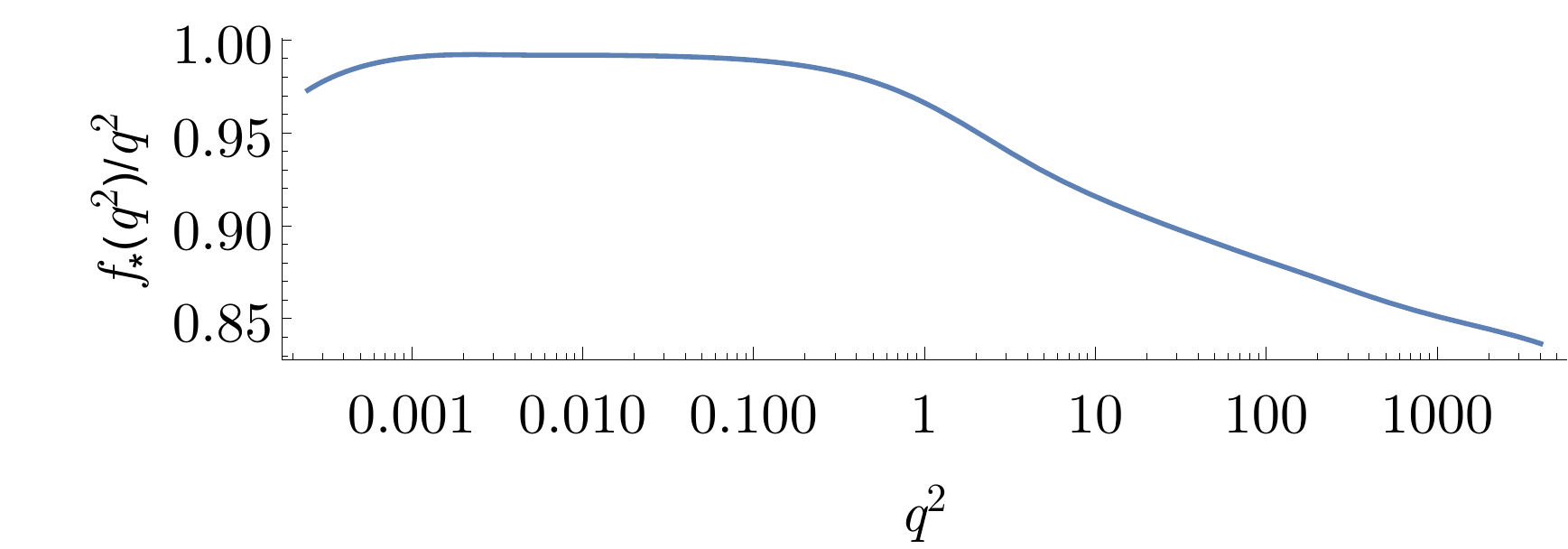}
	\caption{Top: fixed point solution $f_\ast$ (solid yellow). For reference, we have included the line $y=q^2$ (dashed blue). Bottom: ratio $f_\ast(q^2)/q^2$. The function $f_\ast$ only starts to deviate from $y=q^2$ in the regime $q^2 \gtrsim 1$. We note that the dip in the small momentum regime shown in the lower figure constitutes a numerical artefact due to the numerically small but finite gap in our solution.}
	\label{fig:fsol}
\end{figure}

Finally, in order to concretely evaluate the flow equation, we have to choose a regulator function. We choose a regulator of exponential type,
\begin{equation}
			r(z)
	=
			\exp(-c z)
	\,.
\end{equation}
As advocated in \cite{Bosma:2019aiu}, this regulator allows for fast numerical evaluation. The parameter $c$ allows to scan for a range of regulators. In the following, we fix $c=1$.

The described algorithm is executed in Mathematica. Integration over the momentum and angles is done using Mathematica's NIntegrate routine, with GaussKronrodRule as Method option.
We have run the algorithm using $N=50$ coefficients with $32$ digits numerical precision.
In \fref{fig:coeffs}, we plot the absolute values of the coefficients $a_i$. We see that the coefficients converge algebraically up to values of the order $10^{-6}$, which gives an estimate of the error of the numerical result.

\subsubsection{Results}
We will now discuss the properties of the fixed point solution. We find the fixed point values
\begin{equation}
			g_\ast=0.349
	\,,\qquad
			\lambda_\ast=0.300
	\,,\qquad
			\eta_s^\ast=-0.176
	\,,\qquad
			\alpha=0.949
	\,.
\end{equation}

We see that the asymptotic power $\alpha$ is very close to one. Indeed, as is shown in \fref{fig:fsol} the solution only deviates a tiny amount from the Gaussian function $f_{\mathrm{Gaussian}}(z)=z$. However, $f_{\mathrm{Gaussian}}$ is not an exact solution of the fixed point equation. Moreover, we can insert the ansatz $f_{\mathrm{Gaussian}}$ into the right-hand side of \eref{structuremaster} to obtain a one-loop approximation for $f$. Solving the linear differential equation gives however a solution that either has a singularity at zero momentum $q$, or has a root at a positive value of $q$. Both features are undesirable.

Evaluating $f$ directly at zero momentum, we find that $f$ vanishes up to our numerical precision of $\sim10^{-6}$. Thus the fixed point value of the (dimensionless) gap parameter \eref{gapparameter} vanishes,
\be
\mu_\ast = 0 \, . 
\ee
In combination with the positivity properties of $f$ shown in \fref{fig:fsol}, this entails that the scalar propagator at the fixed point corresponds to a single massless scalar degree of freedom. This is also confirmed by checking \eref{eq:zeromomentum}; we find
\begin{equation}\label{eq:L0}
			\mathcal{L}_0	-	\left(1-\frac{1}{2}\eta_s\right)
	=
		-	0.582
	\,.
\end{equation}
Since this is significantly different from zero, we conclude that the fixed point solution is gapless. This supports the conjecture \cite{Eichhorn:2017sok} that the scalar sector of the gravity-matter fixed point possesses a shift symmetry, i.e., it is invariant under $\phi \mapsto \phi + c$ with $c$ being constant.

In order to study the stability of the fixed point, we construct the stability matrix as discussed in \sref{sec:FRG}. At this stage, the following conceptual remark is in order. The relation \eref{Gscalar} (with the regulator $r(z)$ set to zero) shows that the form factor $f$ is closely related to the propagator of the scalar field,
\be
\left(G^{\rm scalar}\right)^{-1}  \propto f(p^2) \propto  \mathcal{Z}^s(p^2) \,  (p^2 + \mu^2) \, . 
\ee
Since $\mu_\ast = 0$, the momentum-dependent wave-function renormalisation $\mathcal{Z}^s(p^2)$ is depicted in the lower panel of \fref{fig:fsol}. Since $\mathcal Z^s_\ast(p^2) > 0$ is positive, it may be absorbed in a field redefinition without affecting the pole structure of the propagator. This suggests to generalise \eref{anomalousdim} to a the momentum-dependent anomalous dimension $\eta_s^\ast(p^2) \equiv - \p_t \ln \mathcal{Z}^s_\ast(p^2)$. It is then natural that there are no stability coefficients associated with deformations of $\mathcal Z^s_\ast(p^2)$.

Owed to this observation, a stability analysis should be limited to the three critical exponents associated with the couplings $\vec{u} = \{g,\lambda,\mu^2\}$ \cite{Christiansen:2014raa}. As an approximation, we perturb the fixed point solution $\vec{u}^\ast = \{g_\ast,\lambda_\ast, f_\ast(z) \}$ by perturbations of the form
\be
\vec{u}  = \vec{u}^\ast + \sum_{I=1}^3 \delta \vec{u}_I \, {\rm exp}(- \theta_I t) \, . 
\ee
Here the $\delta \vec{u}_I$ are the (momentum-independent) eigenvectors associated with the critical exponents $\theta_I$. An exact treatment takes the variation of the scalar wave-function renormalisation into account, which we neglect here. Substituting this ansatz into the flow equations \eref{betag}, \eref{betalambda} and \eref{structuremaster} and linearising in the perturbations then yields
\be\label{c1}
\theta_{1,2} = 1.64 \pm 4.14 \mathbf i 
\, , \qquad 
\theta_3 = 1.16 \, .
\ee
We note that, by the vanishing of the gap parameter at the fixed point, the critical exponent $\theta_3$ is related to \eref{eq:L0} by a factor of minus two, see also \eref{structuremaster}.
Thus the NGFP comes with three relevant directions, acting as a UV attractor for $g, \lambda$ and the gap parameter $\mu$. The complex pair of critical exponents $\theta_{1,2}$ is characteristic for studying gravity and gravity-matter systems in the present setup. It already occurs in the approximation \eref{approx1} where one finds
\be\label{c2}
\theta_{1,2} = 1.64 \pm 4.12 \mathbf i \, . 
\ee
This indicates that our analysis provides the first study of the gravity-matter fixed point identified in \cite{Percacci:2002ie,Dona:2013qba,Meibohm:2015twa,Becker:2017tcx} at the level of self-consistent form factors. Comparing the critical exponents \eref{c1} and \eref{c2} furthermore suggests that the properties of the fixed point are essentially determined by the gravitational contributions \cite{Meibohm:2015twa}. Including a self-consistent form factor in the scalar sector has a very mild effect on the stability coefficients.

%------------------------------------------------------------------
\section{Discussion and conclusion}
\label{sec:discussion}
%------------------------------------------------------------------
In this work we provided a detailed account on momentum-dependent form factors in quantum field theory and their role in the Asymptotic Safety program. From the pioneering works \cite{Christiansen:2014raa, Bosma:2019aiu}, it is already clear that the form factors related to the two-point functions of the fluctuation fields (non-perturbative propagators) hold the key for understanding the structure of spacetime at short distances. Also the fate of spacetime singularities, omnipresent in classical General Relativity, is closely linked to understanding the structure of the graviton propagator at trans-Planckian momenta.

Conceptually, the role of the form factors may be understood as follows. The object carrying the relevant information on the quantum theory is the quantum effective action $\Gamma \equiv \Gamma_{k =0}$. This functional serves as a generating function for all 1PI correlation functions. It is obtained as the endpoint of an RG trajectory where all quantum fluctuations are integrated out. Asymptotic Safety then entails that all couplings $\bar{u}_i(k=0)$ in $\Gamma$ are determined by the fundamental parameters identifying the RG trajectory infinitesimally close to the non-Gaussian fixed point. The map between the fundamental parameters and the $\bar{u}_i(k=0)$ is obtained by solving the flow equation and constructing the complete RG trajectory. This picture entails some profound consequences.

Firstly, the momentum dependence of the couplings $\bar{u}_i(k=0)$, discussed in the introduction, originates from momentum-dependent form factors \emph{evaluated at $k=0$}. A priori, this is conceptually different from the dependence of the couplings $\bar{u}_i(k)$ on the coarse-graining scale $k$ which arises from integrating out quantum fluctuations shell-by-shell in momentum space. In simple cases, the $k$-dependence of the couplings may be identified with their dependence on an external momentum $p$. This is the case if $p$ provides a physical cutoff scale which suppresses the contribution of modes with $p^2 < k^2$ in the case when $k$ is send to zero. The form factors capturing the momentum dependence of the interaction vertices go far beyond this approximation though. In particular, they also cover ``anisotropic'' situations where external momenta differ from each other.

Secondly, as discussed in \sref{sec:struc-funcs-grav}, the cosmological constant and Newton's coupling cannot be promoted to momentum-dependent form factors, since the inclusion of the differential operators leads to surface terms in $\Gamma_k$. Thus $G_{k=0}$ and $\Lambda_{k=0}$ are numbers which are independent of the external momenta in a scattering process. In this way the RG picture is reconciled with the statement that the (renormalised) cosmological constant and Newton's coupling ``do not run'' \cite{Lucat:2018slu}.

The actual computation of the form factors featuring in Asymptotic Safety is still in its infancy. Our goal was to give a detailed (and hopefully accessible) survey of the computational techniques which allow to explore this new research area. We illustrated these techniques based on the simplest example by computing the gravitational corrections to the form factor governing the propagation of a scalar field minimally coupled to the Einstein-Hilbert action. Structurally, the beta functions governing the flow of the scalar propagator, \eref{structuremaster}, already exhibit all the features also expected in the gravitational sector or more complex gravity-matter systems: the scale-dependence is encoded in a non-linear integro-differential equation whose solutions may be found via sophisticated pseudo-spectral methods. 

The main result of this analysis is shown in \fref{fig:fsol}. The result establishes that the scalar kinetic term of the gravity-scalar fixed point previously studied in \cite{Percacci:2002ie,Dona:2013qba,Meibohm:2015twa,Biemans:2017zca,Alkofer:2018fxj,Eichhorn:2018akn} extends to a complete form factor which is well-defined for all momenta. Remarkably, the gravitational corrections to the scalar propagator are rather mild. $f_\ast(p^2)$ essentially follows the classical propagator $f_\ast(p^2) \propto p^2$ with a slightly slower increase at asymptotically large momenta.

The analysis of perturbations around this solution shows that, as expected, the gap parameter of the scalar is a relevant parameter. The scalar anomalous dimension is negative and takes the value $\eta^s_\ast = -0.176$. \Tref{tab.etas} compares this value to results reported in the literature. 
\begin{table}[t!]
	\centering
	\caption{\label{tab.etas} Comparison of the scalar anomalous dimension $\eta^{\rm s}_\ast$ reported in the literature. As an important novel feature, the form factor $f_\ast^{(\phi\phi)}(p^2)$ allows to analyse the asymptotic behaviour of the scalar two-point function (propagator) at large momenta.} 
	\lineup
	\begin{indented}
	\item[]\begin{tabular}{@{}lll}
	\br
	reference & $\eta^{\rm s}_{*}$ & $\Delta$ \\ 	\mr
	\cite{Dona:2013qba} & $ -0.361$ & ---\\
	\cite{Meibohm:2015twa} & $\m0\0\0\0$ & --- \\
		\cite{Becker:2017tcx} & $-0.771$ & --- \\
	this work & $-0.176$ & $1.051$ \\ \br
	\end{tabular}
	\end{indented}
\end{table}
Since the underlying computations use different choices for the field decomposition and the implementation of the regulator function \cite{Dona:2013qba} or the closure of the split symmetry \cite{Meibohm:2015twa} one should not expect a matching of the results beyond the qualitative agreement shown by the table. 

The scaling of the scalar propagator in the large momentum regime follows from the scaling analysis in section \ref{sec:scal-tens-fluc-st-analytic} and is given by
\be
\lim_{p^2 \rightarrow \infty} G^{\rm s}_\ast(p^2) \propto \frac{1}{p^{2\alpha}} \, , \qquad \alpha = 0.949 \, .  
\ee
By performing a Fourier transform to position space, this asymptotic behaviour then governs the short-distance asymptotics of the scalar two-point correlator. Explicitly,
\be
\langle \phi(x) \phi(y) \rangle \simeq \frac{1}{|x-y|^{2\Delta}} \, , \quad \Delta = \frac{1}{2}(d-2\alpha) = 1.051 \, . 
\ee
Notably, this fall-off is compatible with the unitarity bounds on the scaling behaviour of scalar correlators stating that $\Delta \ge \Delta_{\rm min} = 1$ \cite{Rychkov:2016iqz,Simmons-Duffin:2016gjk}.

An important point, which so far has not been addressed in the literature, is the relation between the form factors at the fixed point $f_\ast(\{p_i\})$ and in the quantum effective action $f_{k=0}(\{p_i\})$. Establishing this connection requires solving the $k$-dependent IDE with suitable boundary conditions. In the scalar case analysed in this work this computation would proceed as follows. The first step constructs the eigenperturbations associated with the three relevant directions explicitly. Adding perturbations into the relevant directions then gives different initial conditions for the projected flow equation imposed at asymptotically large values of $k$. The latter is then mapped to the couplings appearing in the quantum effective action $\Gamma_{k=0}$ by solving the IDE \eref{structuremaster}, looking for complete solutions which extend down to $k=0$. While this analysis is crucial for understanding whether the non-Gaussian fixed point is connected to the observed ``low-energy world'', this computation is beyond the scope of the present work.\footnote{For some recent analysis along these lines see \cite{Gubitosi:2018gsl,Saueressig:2019fmx}.}

An important cross-check along these lines may be provided by the effective field theory treatment of gravity \cite{Donoghue:1994dn}, recently reviewed in \cite{Donoghue:2017pgk}. Suppose that instead of tracing the renormalisation group flow to the deep infrared, $k=0$, the solution of the IDE is constructed up to a finite scale $k^2 = \Lambda^2 \lesssim M_{\rm Pl}^2$. Provided that the structure of the effective average action at this scale, $\Gamma_{k=\Lambda}$, resembles the Einstein-Hilbert action, the resulting quantum effective action should resemble the one found in the effective field theory treatment, at least at the perturbative level.

Clearly, understanding the momentum-dependent form factors associated with the non-Gaussian fixed points appearing in gravity and gravity-matter systems and the resulting quantum effective actions constitute formidable computational tasks. Since form factors may be the key towards 
unlocking some of the most fundamental questions in quantum gravity research addressing these challenges may very well be worth the effort.

%-----------------------------------------------------------------------
\ack
%-----------------------------------------------------------------------
We thank D.\ Becker, S.\ Lippoldt, R.\ Percacci, J.M. Pawlowski, T.\ Prokopec, and M.\ Reuter for interesting discussions. This research is supported by the Netherlands Organisation for Scientific Research (NWO) within the Foundation for Fundamental Research on Matter (FOM) grant 13VP12.

\clearpage

\appendix
%------------------------------------------------------------------
\section{Notation and Conventions}
\label{app:notation}
%------------------------------------------------------------------
Symmetrisation and anti-symmetrisation are denoted by round and square brackets
and normalised to unit strength,
\be
H_{\alpha\beta} = H_{(\alpha\beta)} + H_{[\alpha\beta]} \, . 
\ee
The Euclidean spacetime $\cM$ has dimension $d$ and carries a Euclidean metric $g_{\mu\nu}$ with signature $(+,+,\cdots)$. All quantities constructed from the background metric $\gb_{\mu\nu}$ are indicated with a bar.
 The covariant derivative $D_\mu$ contains the Christoffel symbol $\Gamma^\alpha{}_{\mu\nu}$,
\be
D_\mu \, H^\alpha_{\phantom{\alpha}\beta} = \p_\mu H^\alpha{}_\beta + \Gamma^\alpha_{\phantom{\alpha}\mu\sigma} \, H^\sigma_{\phantom{\sigma}\beta} - \Gamma^\sigma_{\phantom{\sigma}\mu\beta} \, H^\alpha_{\phantom{\alpha}\sigma} \, . 
\ee
Throughout the work it is assumed that all fields have suitable fall-off properties so that integration by parts does not generate any surface terms.
The Riemann tensor is defined as
\be
R_{\mu\nu\rho}^{\phantom{\mu\nu\rho}\lambda} \equiv \p_\nu \Gamma^\lambda_{\phantom{\lambda}\mu\rho} - \p_\mu \Gamma^\lambda_{\phantom{\lambda}\nu\rho} + \Gamma^{\sigma}_{\phantom{\sigma}\mu\rho} \Gamma^\lambda_{\phantom{\lambda}\nu\sigma} - \Gamma^\sigma_{\phantom{\sigma}\nu\rho} \Gamma^\lambda_{\phantom{\lambda}\mu\sigma} \, , 
\ee
such that the commutator of two covariant derivatives satisfies
\be
\left[ \, D_\mu \, , \, D_\nu \, \right] H_\lambda = \,  R_{\mu\nu\lambda}^{\phantom{\mu\nu\lambda}\rho} \, H_\rho \, .
\ee
The Ricci tensor and Ricci scalar are $R_{\mu\nu} = R_{\mu\lambda\nu}^{\phantom{\mu\lambda\nu}\lambda} $ and $R = g^{\mu\nu} R_{\mu\nu}$.
The Weyl tensor is the traceless part of the Riemann tensor,
\be\label{Weyldef}
\fl C_{\mu\nu\rho\sigma} = R_{\mu\nu\rho\sigma} - \frac{2}{d-2} \left( g_{\mu[\rho} R_{\sigma]\nu} -  g_{\nu[\rho} R_{\sigma]\mu} \right) + \frac{2}{(d-1)(d-2)} \, R \, g_{\mu[\rho} g_{\sigma]\nu} \, . 
\ee
The Riemann tensor satisfies the first and second Bianchi identity,
\be\label{Bianchi12}
R_{\mu[\nu\rho\sigma]} = 0 \, , \quad D_{[\alpha} R_{\mu\nu]\rho\sigma} = 0 \, .
\ee
The latter also implies that
\be\label{Bianchi2contracted}
D^\alpha R_{\alpha\beta\mu\nu} = 2 D_{[\mu} R_{\nu]\beta}  \, , \qquad D^\nu R_{\mu\nu} = \frac{1}{2} D_\mu R \, . 
\ee
The (positive definite) Laplacian is denoted by
\be
\Delta \equiv - g^{\mu \nu} D_\mu D_\nu \equiv - D^2 \, , 
\ee
and we use the symbol $\Box$ for the Laplacian in flat space,
\be
\Box \equiv - \delta^{\mu\nu} \p_\mu \p_\nu \, .
\ee

Action functionals contain strings of fields built out of the fundamental fields themselves or in form of curvature tensors. The projection of an action to a string of fields is denoted by the action monomial followed by a vertical line carrying the string of fields onto which the action is projected as a subscript. For instance, the projection of the volume term to the string given by one power of the metric fluctuation is denoted by
\be
\left. \int \rmd^dx \sqrt{g} \, \right|_h = \frac{1}{2} \int \rmd^dx \sqrt{\gb} \, h \, . 
\ee
 Derivatives acting on the $i$-th field in the string carry the number of the field as a subscript, i.e.,
\be
\int \rmd^dx \sqrt{g} \, (\Delta_1 \Delta_2 \Delta_3) (R_1 R_2 R_3) = \int \rmd^dx \sqrt{g} (\Delta R_1) (\Delta R_2) (\Delta R_3) \, . 
\ee
Notably differential operators with different subscripts commute since they are acting on different fields.

In many cases the computation can be simplified by adopting flat Euclidean space as background. In this case it is convenient to switch to momentum space and work with the Fourier-transformed fields,
\be\label{Fourier-transform}
\phi(p) = \int \! \rmd^{d}x \, \phi(x) \, e^{-\mathbf{i}px} \, ,
\ee
where we use the same symbol for the fields in position and momentum space. In this case the form factors depend on the field's momenta. We adopt the convention that all momenta are incoming, so that momentum conservation at an $n$-point vertex implies
\be\label{momentumconservation}
\sum_{i=1}^n p_i = 0 \, .
\ee
Starting from an action functional, the resulting interaction vertices are obtained by taking suitable variations with respect to the (fluctuation) fields. This automatically leads to a symmetrisation in the tensor and momentum structures, e.g.
\begin{eqnarray}
\frac{\delta}{\delta \phi(p_4)} \frac{\delta}{\delta \phi(p_3)} \frac{\delta}{\delta h_{\rho\sigma}(p_2)} \frac{\delta}{\delta h_{\mu\nu}(p_1)} h_{\alpha\beta}(q_1) h_{\gamma\delta}(q_2) \phi(q_3) \phi(q_4) \nonumber \\
= \bigg[ \mathbbm 1_{\alpha\beta}^{\phantom{\alpha\beta}\mu\nu} \mathbbm 1_{\gamma\delta}^{\phantom{\gamma\delta}\rho\sigma} \delta(q_1-p_1) \delta(q_2-p_2) + \mathbbm 1_{\alpha\beta}^{\phantom{\alpha\beta}\rho\sigma} \mathbbm 1_{\gamma\delta}^{\phantom{\gamma\delta}\mu\nu} \delta(q_1-p_2) \delta(q_2-p_1) \bigg] \times \nonumber \\
\qquad \bigg[ \delta(q_3 - p_3) \delta(q_4 - p_4) + \delta(q_3 - p_4) \delta(q_4 - p_3) \bigg] \, ,
\label{fctvariation}
\end{eqnarray}
for the $(hh\phi\phi)$-vertex. Here, the identity on the space of symmetric $d\times d$ matrices is
\begin{equation}
{\mathbbm 1_{\alpha\beta}}^{\mu\nu} \equiv \frac{1}{2} \bigg( \delta_\alpha^\mu \delta_\beta^\nu + \delta_\alpha^\nu \delta_\beta^\mu \bigg) \, 
\end{equation}
and $\delta(x)$ is the usual Dirac delta in $d$ dimensions.
\newpage
%------------------------------------------------------------------
\section{Expansion in metric fluctuations}
\label{app:expansion}
%------------------------------------------------------------------
Constructing solutions of the flow equation requires taking functional derivatives with respect to the metric fluctuations. The expansions of various basic quantities entering the computations in this work are listed in Table \ref{tab.expansion}. Notably, the complexity of the tensor structures entering the computations increases rapidly. The prototypical example is the vertex $\Delta C^{\mu\nu\rho\sigma}|_{hh}$ which arises from the form factor \eref{structure2a}. It is then economical to construct the variations of the required invariants using suitable computer algebra packages like xAct \cite{xActwebpage} and its extensions \cite{2007CoPhC.177..640M,Brizuela:2008ra,2008CoPhC.179..597M,2014CoPhC.185.1719N}.
\begin{table}[h!]
\centering
	\caption{\label{tab.expansion} Expansion of selected covariant objects in terms of metric fluctuations $g_{\mu \nu} = \gb_{\mu \nu} + h_{\mu \nu}$ at a fixed order in $h_{\mu\nu}$. Indices are raised with $\gb^{\mu\nu}$ and the trace of the fluctuation field is denoted by $h \equiv \gb^{\mu\nu} h_{\mu\nu}$. The last two entries exemplify expansions that are typical when computing momentum-dependent form factors in a fluctuating spacetime.}
	\renewcommand{\arraystretch}{1.4}
	\begin{indented}\item[]
	\begin{tabular*}{\linewidth}{lcl@{\extracolsep{\fill}}}
		\br
		$g_{\mu\nu}|_h$ & $=$ & $h_{\mu\nu}$ \\ 
		$g^{\mu\nu}|_h$ & $=$ &  $-h^{\mu\nu}$ \\ 
		$\Gamma^\lambda_{\phantom{\lambda}\mu\nu}|_h$ & $=$ & $\frac{1}{2}  \gb^{\lambda\sigma}  \left( \Db_\mu h_{\nu\sigma} + \Db_\nu h_{\mu\sigma} - \Db_\sigma h_{\mu\nu} \right) $ \\
		$\left.\frac{\sqrt{g}}{\sqrt{\gb}}\right|_h$ & $=$ &  $\frac{1}{2} h$  \\ 
%		%
		$R_{\mu\nu}|_h$ & $=$ & $ \Rb_{\sigma(\mu}^{\phantom{)}} h_{\nu)}^{\phantom{\nu)}\sigma} 
		+  \Rb_{\sigma\mu\nu\lambda} h^{\sigma\lambda} 
		+ \Db_{(\mu}^{\phantom{)}} \Db_{|\sigma|}^{\phantom{)}} h_{\nu)}^{\phantom{\nu)}\sigma} - \frac{1}{2} \Db_\mu \Db_\nu h - \frac{1}{2} \Db^2 h_{\mu\nu} $ \\
%		%
		$R|_h$ & $=$ &  $-\Rb^{\mu\nu} h_{\mu\nu} + \Db_\alpha \Db_\beta h^{\alpha\beta} - \Db^2 h $ \\ 
		$\left.\frac{\sqrt{g}}{\sqrt{\gb}}\right|_{hh}$ & $=$ &  $\frac{1}{8} h^2 - \frac{1}{4} h^{\mu\nu} h_{\mu\nu}$ \\ 
%		%
		$R|_{hh}$ & $=$ & $\Rb_{\alpha\gamma\beta\delta} h^{\alpha\beta} h^{\gamma\delta} + h^{\alpha\beta} \Db^2 h_{\alpha\beta} + h^{\alpha\beta} \Db_\alpha \Db_\beta h - 2 h^{\alpha\beta} \Db_\beta \Db^\gamma h_{\alpha\gamma} $  \\
%		%
		&& $-\frac{1}{4} \left(\Db^\alpha h\right) \Db_\alpha h + \frac{3}{4} \left( \Db^\gamma h^{\alpha\beta} \right) \Db_\gamma h_{\alpha\beta} - \left(\Db_\alpha h^{\alpha\beta} \right) \Db^\gamma h_{\beta\gamma} $ \\
%		%
		&& $-\frac{1}{2} \left( \Db^\alpha h^{\beta\gamma} \right) \Db_\gamma h_{\alpha\beta} + \left( \Db^\alpha h \right) \Db^\beta h_{\alpha\beta}$ \\ \mr
		$(\Delta \phi)|_h$ & $=$ & $\bigg[h_{\mu\nu}\bar D^\mu \bar D^\nu + (\bar D^\mu h_{\mu\nu}) \bar D^\nu - \frac{1}{2} (\bar D_\mu h) \bar D^\mu\bigg]\phi$ \\ 
%		 %
		 	$(\Delta \phi)|_{hh}$ & $=$ & $\bigg[-h_\mu^{\phantom{\mu}\alpha} h_{\alpha\nu} \bar D^\mu \bar D^\nu - h^{\alpha\beta}(\bar D_\beta h_{\alpha\mu}) \bar D^\mu - h_\mu^{\phantom{\mu}\beta} (\bar D^\gamma h_{\beta\gamma}) \bar D^\mu$  \\
	 	&& $+ \frac{1}{2} h^{\alpha\beta} (\bar D_\mu h_{\alpha\beta}) \bar D^\mu + \frac{1}{2} h^{\mu\nu} (\bar D_\mu h)\bar D_\nu \bigg]\phi$ \\ 
		 \br
	\end{tabular*}
	\end{indented}
\end{table}

%------------------------------------------------------------------
\newpage
\section{Varying functions of Laplacians - useful formulas}
\label{app:laplacian}
%------------------------------------------------------------------
In this appendix we set up the machinery to expand of functions $f(\Delta)$ of the Laplacian to a given order in the fluctuation field $h_{\mu\nu}$. The objective is to \emph{retain all covariant derivatives}, which is central for computing the momentum dependence of the form factors. We start by introducing the concept of multi-commutators and their properties in \Sref{sec:C.1} before discussing the expansion of $f(\Delta)$ in \Sref{sec:C.2}.  
%------------------------------------------------------------------
\subsection{Multi-commutators and their combinatorical identities}
\label{sec:C.1}
%------------------------------------------------------------------
Let $Q,Y,X,Z$ denote some (differential) operators. The multi-commutator is then defined recursively as
\be
{}[X, Y]_l \equiv [X, [X, Y]_{l-1} ] \, , \qquad {}[X, Y]_0 = Y \, , \quad l \ge 0 \in \mathbb{N}. 
\ee
For $l=1$ it reduces to the standard commutator
\be
[X, Y]_1 = [X,Y] = XY-YX \, .
\ee
The multi-commutator is linear in its second argument,
\be
[X, Y+Z]_l = [X, Y]_l + [X, Z]_l \, , 
\ee
and, for a constant parameter $s$, obeys the scaling relations
\be\label{scalingprop}
{}[s X, Y]_l = s^l \, {}[X, Y]_l \, , \quad {}[X, s Y]_l = s \, {}[X, Y]_l \, . 
\ee
A multi-commutator containing a product of operators in the second argument can be expressed as a finite sum
\be
{}[X, YZ]_m = \sum_{l=0}^m  {m \choose l} \, \left[X,Y\right]_l  \, \left[ X,Z\right]_{m-l} \, .
\ee
For $X=Z=\Delta$ this entails the useful identity
\be
\left[ \Delta, Y \Delta\right]_l = \left[ \Delta, Y \right]_l \Delta \, . 
\ee

The multi-commutator allows to give exact expressions for commuting differential operators. In particular one can proof by induction that
\begin{equation}\label{eq.C7}
Q^m Y = \sum_{l=0}^m {m \choose l} [Q,Y]_l Q^{m-l} \, , 
\end{equation}
and
\begin{equation}\label{eq.C8}
Y Q^m = \sum_{l=0}^m {m \choose l} (-1)^l Q^{m-l} [Q,Y]_l \, .
\end{equation}
Finally, we note that when integrated over spacetime, multi-commutators may be resolved by employing the identity
\begin{equation}\label{undomulti}
\fl \int \rmd^dx \sqrt{g} \, Y \left[\Delta, Z \right]_m \, X = \int \rmd^dx \sqrt{g} \, \sum_{l=0}^m {m \choose l} (-1)^l \left( \Delta^{m-l} Y \right) Z \left(\Delta^l X \right) \, ,
\end{equation}
which is again proven by induction.
%------------------------------------------------------------------
\subsection{Expanding functions of the Laplacian in terms of fluctuation fields}
\label{sec:C.2}
%------------------------------------------------------------------
We are now in the position to evaluate the expansion of $f(\Delta)$ in terms of fluctuations $h_{\mu\nu}$ around a fixed background metric $\gb_{\mu\nu}$. In the first step we express $f(\Delta)$ in terms of the inverse Laplace transform $\tilde f(s)$,
\begin{equation}\label{LaplaceTrafo}
f(\Delta) = \int_0^\infty \rmd s \, \tilde f(s) \, \rme^{-s\Delta} \, ,
\end{equation}
which we always assume to exist. This covers in particular logarithms of the Laplacian which can be represented as  \cite{Donoghue:2015nba},
\begin{equation}
\ln(\Delta) = \int_0^\infty \rmd s \,  \frac{\rme^{-s}-\rme^{-s\Delta}}{s} \, .
\end{equation}
At this stage, the problem of carrying out the $h$-expansion simplifies to expanding the exponential function and subsequently undoing the Laplace transform.

We then note that the Laplacian $\Delta = - g^{\mu\nu}D_\mu D_\nu$ admits an expansion
\begin{equation}\label{deltaexp}
\Delta = \bar \Delta + \mathbbm d_1 + \mathbbm d_2 + \dots \, ,
\end{equation}
where $\bar\Delta$ is the Laplacian constructed from the background metric and the expansion coefficients $\mathbbm d_m$ contain $m$ powers of the metric fluctuation $h_{\mu\nu}$. The $\mathbbm d_m$ depend on the tensor structure on which the Laplacian acts. For example, for a scalar $\phi$, we have
\begin{eqnarray}
\mathbbm d_1^\phi &= h_{\mu\nu}\bar D^\mu \bar D^\nu + (\bar D^\mu h_{\mu\nu}) \bar D^\nu - \frac{1}{2} (\bar D_\alpha h) \bar D^\alpha \, , \label{eq:d1scalar} \\
\mathbbm d_2^\phi &= -{h_\mu}^\alpha h_{\alpha\nu} \bar D^\mu \bar D^\nu - h^{\alpha\beta}(\bar D_\beta h_{\alpha\mu}) \bar D^\mu - {h_\mu}^\beta (\bar D^\gamma h_{\beta\gamma}) \bar D^\mu \nonumber \\
&\qquad + \frac{1}{2} h^{\alpha\beta} (\bar D_\mu h_{\alpha\beta}) \bar D^\mu + \frac{1}{2} h^{\mu\nu} (\bar D_\mu h)\bar D_\nu \, . \label{eq:d2scalar}
\end{eqnarray}

The next step substitutes the expansion \eref{deltaexp} into the exponential and subsequently expands in the fluctuation field. At this stage we first note the auxiliary identity
\begin{eqnarray}
\frac{d}{d\epsilon} \rme^{X+\epsilon Y} &=  V(\epsilon; X,Y) \, \rme^{X+\epsilon Y} = \rme^{X+\epsilon Y} \, \tilde V(\epsilon; X,Y) \, .
\end{eqnarray}
where
\begin{eqnarray}
 V(\epsilon; X,Y) &= \sum_{j=0}^\infty \frac{1}{(j+1)!} [X+\epsilon Y,Y]_j \, ,  \\
\tilde V(\epsilon; X,Y) &= \sum_{j=0}^\infty \frac{(-1)^j}{(j+1)!} [X+\epsilon Y,Y]_j \, .
\end{eqnarray}
The expressions for $V(\epsilon; X,Y)$ and $\tilde V(\epsilon; X,Y)$ thereby follow from expanding the exponential in its Taylor series, taking the derivative with respect to $\epsilon$ and a subsequent reordering of terms employing the identities \eref{eq.C7} and \eref{eq.C8}. This result can then be used to construct the expansion of $\rme^{X+\epsilon Y}$ in $\epsilon$ either bringing the exponential factor to the left
\be
\fl \rme^{X+\epsilon Y} = \rme^X \left[ 1 + \epsilon \tilde V(0;X,Y) + \frac{\epsilon^2}{2} \left( \tilde V(0;X,Y)^2 + \tilde V_\epsilon(0;X,Y) \right) + \mathcal O(\epsilon^3) \right] \, ,
\ee
or to the right
\be
\fl \rme^{X+\epsilon Y} = \left[ 1 + \epsilon V(0;X,Y) + \frac{\epsilon^2}{2} \left( V(0;X,Y)^2 + V_\epsilon(0;X,Y) \right) + \mathcal O(\epsilon^3) \right] \rme^X \, .
\ee
Here, the subscript $\epsilon$ indicates a derivative w.r.t.\ $\epsilon$ before setting $\epsilon$ to zero. A straightforward calculation shows that
\begin{equation}
V_\epsilon(0;X,Y) = \sum_{j=0}^\infty \sum_{k=1}^\infty \frac{1}{(k+j+2)!}[X,[Y,[X,Y]_k]]_j \, ,
\end{equation}
and
\begin{equation}
\tilde V_\epsilon(0;X,Y) = \sum_{j=0}^\infty \sum_{k=1}^\infty \frac{(-1)^{k+j+1}}{(k+j+2)!}[X,[Y,[X,Y]_k]]_j \, .
\end{equation}
Replacing the operators $X$ and $Y$ by the background Laplacian $\bar \Delta$ and the expansion coefficients $\mathbbm d_m$ then allows to extract the required powers of the fluctuation field from the exponential $\rme^{-s \Delta}$. We stress that this expansion is exact in the sense that there is no approximation on the momentum structure, i.e., all derivatives acting on fields are retained.

We conclude our derivation with a summary of the algorithm described above:
\begin{enumerate}
	\item rewrite the function as a Laplace transform,
	\item calculate the coefficients $\mathbbm d_i$ defined in \eref{deltaexp} up to the required order,
	\item rescale the metric fluctuation $h_{\mu\nu}$ by a control parameter $\epsilon$,
	\item choose whether to sort the exponential of the background Laplacian to the left or the right of the expansion,
	\item use the corresponding expansion formula and set $\epsilon$ to $1$ after truncating the series at the desired order,
	\item undo the Laplace transform.
\end{enumerate}
Depending on the concrete situation, once the expansion is done one can do the sums over the multi-commutators using the scaling properties \eref{scalingprop} together with identities of the type \eref{undomulti} and perform the Laplace transform. If this is not possible, one can nevertheless handle all the expressions as they are, and perform the sums and the transform at the very end, after the functional trace has been calculated.

\newpage

%------------------------------------------------------------------
\section{\texorpdfstring{The four-point vertex $\Gamma^{(hh\phi\phi)}$}{The four-point vertex Gamma-h-h-phi-phi}}
\label{app:4ptvertex}
%------------------------------------------------------------------

In this appendix we discuss the general structure of the $(hh\phi\phi)$-vertex. Most of the discussion straightforwardly extends to general four-point vertices. We first note that any four-point form factor has six independent variables, which we take as the squares of three momenta and the scalar products between them,
\begin{equation}
 f^{(hh\phi\phi)}_{\mathcal T} = f^{(hh\phi\phi)}_{\mathcal T}(p_1^2,p_2^2,p_3^2,y_{12},y_{13},y_{23}) \, .
\end{equation}
In this,
\begin{equation}
 p_{i\mu} p_j^\mu = y_{ij} \, ,
\end{equation}
denotes the scalar product between two different momenta. All other combinations can be related to these variables by momentum conservation:
\begin{eqnarray}
 p_4^2 &= p_1^2 + p_2^2 + p_3^2 + 2y_{12} + 2y_{13} + 2y_{23} \, , \\
 y_{14} &= -p_1^2 - y_{12} - y_{13} \, , \\
 y_{24} &= -p_2^2 - y_{12} - y_{23} \, , \\
 y_{34} &= -p_3^2 - y_{13} - y_{23} \, .
\end{eqnarray}
If one wants to resolve any four-point function numerically, it is useful to parameterise the (cosine of the) angles instead of the full scalar product to get a fixed domain in all variables.

Note that starting from the four-point function, the covariantisation to curved space is non-trivial. For the three-point function, no ordering ambiguity exists since the different Laplacians commute as they act on different objects. Here this is no longer the case since $\Delta_1$ and $D_{1\mu}D_2^\mu$, etc., do no longer commute.

Now we have to find an operator basis. There are five different types of tensor structures, depending on the number of derivatives contracted with the graviton fluctuations. To write down a basis, we introduce a short hand notation on the basis of lexicographic ordering of indices. In general, a tensor structure looks like
\begin{equation}
 \mathcal T^{\alpha\beta\gamma\delta} h_{\alpha\beta}(p_1) h_{\gamma\delta}(p_2) \phi(p_3) \phi(p_4) \, ,
\end{equation}
where $\mathcal{T}$ can consist of either the background metric or derivatives acting on either field. We will choose the convention that all derivatives acting on $\phi(p_4)$ will be integrated by parts. Then, there are the following 59 form factors. First, those where $\mathcal{T}$ only contains the metric:
\begin{eqnarray}
 \mathcal T^{\alpha\beta\gamma\delta} &= \frac{1}{2}\left( \bar g^{\alpha\gamma} \bar g^{\beta\delta} + \bar g^{\alpha\delta} \bar g^{\beta\gamma} \right) \, \leftrightarrow \, f^{(hh\phi\phi)}_{(\mathbbm 1)} \, , \nonumber \\
 \mathcal T^{\alpha\beta\gamma\delta} &= \frac{1}{d} \bar g^{\alpha\beta} \bar g^{\gamma\delta} \, \leftrightarrow \, f^{(hh\phi\phi)}_{(\bar g \bar g)} \, .
\end{eqnarray}
There are three different types of structures with one metric and two derivatives in $\mathcal{T}$. Using lexicographic ordering, we denote the six functions of type 1 by
\begin{equation}
 \mathcal T^{\alpha\beta\gamma\delta} = \bar g^{\alpha\beta} \partial_i^\gamma \partial_j^\delta \, \leftrightarrow \, f^{(hh\phi\phi)}_{(\bar g \, i j)} \, ,
\end{equation}
where $i\in\{1,2,3\}$ and $j\in\{1,2,3\}, j\geq i$, indicate which field the derivatives acts upon. For example, the expression $(\bar g 1 3)$ denotes the term
\begin{equation}
 f^{(hh\phi\phi)}_{(\bar g 1 3)}(p_1^2,p_2^2,p_3^2,y_{12},y_{13},y_{23}) (\partial^\gamma h(p_1)) h_{\gamma\delta}(p_2) (\partial^\delta \phi(p_3)) \phi(p_4) \, .
\end{equation}
In general independent of this (except for exceptional momentum configurations) are the six terms
\begin{equation}
 \mathcal T^{\alpha\beta\gamma\delta} = \bar g^{\gamma\delta} \partial_i^\alpha \partial_j^\beta \, \leftrightarrow \, f^{(hh\phi\phi)}_{(i j \, \bar g)} \, ,
\end{equation}
with the same set of possibilities for $i$ and $j$ as above. The third type is
\begin{equation}
 \mathcal T^{\alpha\beta\gamma\delta} = \bar g^{\beta\delta} \partial_i^\alpha \partial_j^\gamma \, \leftrightarrow \, f^{(hh\phi\phi)}_{(i \, \bar g \, j)} \, ,
\end{equation}
where both $i$ and $j$ are in $\{1,2,3\}$, giving rise to nine combinations. Finally, there are 36 terms without metrics,
\begin{equation}
 \mathcal T^{\alpha\beta\gamma\delta} = \partial_i^\alpha \partial_j^\beta \partial_k^\gamma \partial_l^\delta \, \leftrightarrow \, f^{(hh\phi\phi)}_{(ijkl)} \, ,
\end{equation}
where again due to symmetry reasons, $i \leq j$ and $k \leq l$.

Performing a Fourier transform and using \eref{fctvariation}, we have
\begin{eqnarray}
 \Gamma^{(hh\phi\phi)\mu\nu\rho\sigma}(p_1,p_2,p_3,p_4) \nonumber \\
 = \int_{q_1\cdots q_4} \sum_{\mathcal T} \mathcal T^{\alpha\beta\gamma\delta} f^{(hh\phi\phi)}_{\mathcal T}(q_1^2,q_2^2,q_3^2,q_{1\mu}q_2^\mu,q_{1\mu}q_3^\mu,q_{2\mu}q_3^\mu) \times \nonumber \\
 \qquad \bigg[ \delta(q_3 - p_3) \delta(q_4 - p_4) + \delta(q_3 - p_4) \delta(q_4 - p_3) \bigg] \times \\
 \qquad \bigg[ {\mathbbm 1_{\alpha\beta}}^{\mu\nu}{\mathbbm 1_{\gamma\delta}}^{\rho\sigma} \delta(q_1-p_1) \delta(q_2-p_2) + {\mathbbm 1_{\alpha\beta}}^{\rho\sigma}{\mathbbm 1_{\gamma\delta}}^{\mu\nu} \delta(q_1-p_2) \delta(q_2-p_1) \bigg] \, . \nonumber
\end{eqnarray}
It is understood that all occurrences of $p_4$ are replaced by $-p_1-p_2-p_3$.

For the derivation of the flow equation for the kinetic term of the scalar field, we need to derive these 59 form factors from our single-metric ansatz. To lighten the notation, in the following we will suppress the superscript $(\phi\phi)$ on $f$. The expansion of our ansatz in metric fluctuations reads
\begin{eqnarray}
 \fl &\frac{1}{2} \int \rmd^dx \sqrt{g} \, \phi f(\Delta) \phi = \frac{1}{2} \int \rmd^dx \int_0^\infty \rmd s \, \tilde f(s) \, \sqrt{g} \, \phi e^{-s\Delta} \phi \nonumber \\
 \fl &\simeq \frac{1}{2} \int \rmd^dx \int_0^\infty \rmd s \, \tilde f(s) \, \sqrt{\bar g} \, \left[ 1 + \frac{1}{2} h + \frac{1}{8} h^2 - \frac{1}{4} h_{\mu\nu} h^{\mu\nu} \right] \times \\
 \fl &\quad \phi \left[ 1 + V(0;-s\bar\Delta,-s\mathbbm d_1-s\mathbbm d_2) + \frac{1}{2} \left( V(0;-s\bar\Delta,-s\mathbbm d_1)^2 + V_\epsilon(0;-s\bar\Delta,-s\mathbbm d_1) \right) \right] e^{-s\bar\Delta} \phi \, . \nonumber
\end{eqnarray}
We split the contributions to the four-point function into pieces:
\begin{eqnarray}
 \fl &\bullet \frac{1}{8} \int \rmd^dx \sqrt{\bar g} \, \left[ \frac{1}{2} h^2 - h_{\mu\nu} h^{\mu\nu} \right] \phi f(\bar\Delta) \phi \, , \\
 \fl &\bullet \frac{1}{4} \int \rmd^dx \sqrt{\bar g} \int_0^\infty \rmd s \, \tilde f(s) \, h \, \phi \sum_{j\geq 0} \frac{1}{(j+1)!} \left[ -s\bar\Delta, -s\mathbbm d_1 \right]_j e^{-s\bar\Delta} \phi \, , \\
 \fl &\bullet \frac{1}{2} \int \rmd^dx \sqrt{\bar g} \int_0^\infty \rmd s \, \tilde f(s) \, \phi \sum_{j\geq 0} \frac{1}{(j+1)!} \left[ -s\bar\Delta, -s\mathbbm d_2 \right]_j e^{-s\bar\Delta} \phi \, , \\
 \fl &\bullet \frac{1}{4} \int \rmd^dx \sqrt{\bar g} \int_0^\infty \rmd s \, \tilde f(s) \, \phi \sum_{j\geq 0} \frac{1}{(j+1)!} \left[ -s\bar\Delta, -s\mathbbm d_1 \right]_j \times \nonumber \\
 \fl & \qquad\qquad\qquad\qquad\qquad\qquad\qquad\qquad \sum_{l\geq 0} \frac{1}{(l+1)!} \left[ -s\bar\Delta, -s\mathbbm d_1 \right]_l e^{-s\bar\Delta} \phi \, , \\
 \fl &\bullet \frac{1}{4} \int \rmd^dx \sqrt{\bar g} \int_0^\infty \rmd s \, \tilde f(s) \, \phi \sum_{j\geq0} \sum_{k\geq1} \frac{1}{(j+k+2)!} \times \nonumber \\
 \fl & \qquad\qquad\qquad\qquad\qquad\qquad\qquad\qquad \left[ -s\bar\Delta, \left[ -s\mathbbm d_1, \left[ -s\bar \Delta, -s\mathbbm d_1 \right]_k \right] \right]_j e^{-s\bar\Delta} \phi \, .
\end{eqnarray}
The $\mathbbm d_i$ are the ones obtained from the scalar Laplacian. We calculate the general vertex functions in the basis as above. The first term gives contributions
\begin{eqnarray}
 \frac{1}{16} f(p_3^2) &\addsto f^{(hh\phi\phi)}_{(\bar g \bar g)} \, , \nonumber \\
 -\frac{1}{8} f(p_3^2) &\addsto f^{(hh\phi\phi)}_{(\mathbbm 1)} \, .
\end{eqnarray}
Here and in the following, the sign $\addsto$ signals that the term on the left contributes to the form factor(s) on the right. For the second term we calculate
\begin{eqnarray}
 \fl &\frac{1}{4} \int \rmd^dx \sqrt{\bar g} \int_0^\infty \rmd s \, \tilde f(s) \, h \, \phi \sum_{j\geq 0} \frac{(-s)^{j+1}}{(j+1)!} \left[ \bar\Delta, \mathbbm d_1 \right]_j e^{-s\bar\Delta} \phi \nonumber \\
 \fl &=\frac{1}{4} \int \rmd^dx \sqrt{\bar g} \int_0^\infty \rmd s \, \tilde f(s) \, h \, \phi \sum_{j\geq 0} \frac{(-s)^{j+1}}{(j+1)!} (p_1^2+2y_{13})^j \mathbbm d_1 e^{-s p_3^2} \phi \nonumber \\
 \fl &=\frac{1}{4} \int \rmd^dx \sqrt{\bar g} \int_0^\infty \rmd s \, \tilde f(s) \sum_{j\geq 0} \frac{(-s)^{j+1}}{(j+1)!} (p_1^2+2y_{13})^j e^{-s p_3^2} \times \nonumber \\
 \fl &\qquad\qquad\qquad\qquad\qquad\qquad\qquad\qquad \left[ \partial_3^\alpha \partial_3^\beta + \partial_1^\alpha \partial_3^\beta + \frac{1}{2} \bar g^{\alpha\beta} y_{13} \right] h_{\alpha\beta} h \phi \phi  \, .
\end{eqnarray}
Here we chose the momentum of the $h$ in the $\mathbbm d_1$ as $p_1$ and that of the $\phi$ to the right as $p_3$. Such a choice is not a problem since the variation of the vertex gives the correct symmetrisation automatically.
For non-exceptional momenta (which is the case for our tadpole diagram), we can further do the sum, thus the contribution of this term is
\begin{eqnarray}
 \frac{1}{4} \frac{f(p_1^2+2y_{13}+p_3^2)-f(p_3^2)}{p_1^2+2y_{13}} &\addsto f^{(hh\phi\phi)}_{(33\bar g)}, f^{(hh\phi\phi)}_{(13\bar g)} \, , \nonumber \\
 \frac{1}{8} y_{13} \frac{f(p_1^2+2y_{13}+p_3^2)-f(p_3^2)}{p_1^2+2y_{13}} &\addsto f^{(hh\phi\phi)}_{(\bar g \bar g)} \, .
\end{eqnarray}
For the exceptional momentum configuration $p_1^2+2y_{13}=0$, the finite difference goes over to a derivative,
\begin{equation}
 \lim_{p_1^2+2y_{13}\to0} \frac{f(p_1^2+2y_{13}+p_3^2)-f(p_3^2)}{p_1^2+2y_{13}} = f'(p_3^2) \, ,
\end{equation}
which can also directly be verified from the sum representation above, where all terms with $j>0$ vanish, and the Laplace transform can be carried out trivially. As a general strategy we propose to insert the sum representation which involves the inverse Laplace transform of $f$ as above into any given diagram, and only do the sum and back-transform afterwards. This correctly accounts for potentially exceptional momentum configurations.

We continue with the next term,
\begin{eqnarray}
 \fl &\frac{1}{2} \int \rmd^dx \sqrt{\bar g} \int_0^\infty \rmd s \, \tilde f(s) \, \phi \sum_{j\geq 0} \frac{(-s)^{j+1}}{(j+1)!} \left[ \bar\Delta, \mathbbm d_2 \right]_j e^{-s\bar\Delta} \phi \nonumber \\
 \fl &= \frac{1}{2} \int \rmd^dx \sqrt{\bar g} \int_0^\infty \rmd s \, \tilde f(s) \, \phi \sum_{j\geq 0} \frac{(-s)^{j+1}}{(j+1)!} (p_1^2+p_2^2+2y_{12}+2y_{13}+2y_{23})^j \mathbbm d_2 e^{-s p_3^2} \phi \, .
\end{eqnarray}
For this we need
\begin{equation}
 \fl \phi \, \mathbbm d_2 \phi = \bigg[ -\bar g^{\beta\delta} \partial_3^\alpha \partial_3^\gamma - \bar g^{\beta\delta} \partial_3^\alpha \partial_1^\gamma - \bar g^{\beta\delta} \partial_1^\alpha \partial_3^\gamma - \frac{1}{2} y_{13} \bar g^{\alpha\gamma} \bar g^{\beta\delta} + \frac{1}{2} \bar g^{\gamma\delta} \partial_2^\alpha \partial_3^\beta \bigg] h_{\alpha\beta} h_{\gamma\delta} \phi \phi \, .
\end{equation}
In the tadpole diagram this contribution enters with exceptional momenta, we thus follow the strategy advertised above.
Denoting
\begin{equation}
 \mathfrak Z = \frac{1}{2} \int_0^\infty \rmd s \, \tilde f(s) \, \sum_{j\geq 0} \frac{(-s)^{j+1}}{(j+1)!} (p_1^2+p_2^2+2y_{12}+2y_{13}+2y_{23})^j e^{-s p_3^2} \, ,
\end{equation}
the contributions are
\begin{eqnarray}
 -\mathfrak Z &\addsto f^{(hh\phi\phi)}_{(3\bar g 3)}, f^{(hh\phi\phi)}_{(3\bar g 1)}, f^{(hh\phi\phi)}_{(1\bar g 3)} \, , \nonumber \\
 -\frac{1}{2} y_{13} \mathfrak Z &\addsto f^{(hh\phi\phi)}_{(\mathbbm 1)} \, , \nonumber \\
 \frac{1}{2} \mathfrak Z &\addsto f^{(hh\phi\phi)}_{(23\bar g)} \, .
\end{eqnarray}

For the fourth term,
\begin{eqnarray}
 \fl &\frac{1}{4} \int \rmd^dx \sqrt{\bar g} \int_0^\infty \rmd s \, \tilde f(s) \, \phi \sum_{j\geq 0} \frac{1}{(j+1)!} \left[ -s\bar\Delta, -s\mathbbm d_1 \right]_j  \sum_{l\geq 0} \frac{1}{(l+1)!} \left[ -s\bar\Delta, -s\mathbbm d_1 \right]_l e^{-s\bar\Delta} \phi \nonumber \\
 \fl &= \frac{1}{4} \int \rmd^dx \sqrt{\bar g} \int_0^\infty \rmd s \, \tilde f(s) \, \phi \sum_{j\geq 0} \sum_{l\geq 0} \frac{(-s)^{j+l+2}}{(j+1)!(l+1)!} \times \nonumber \\
 \fl & \qquad\qquad\qquad\qquad\qquad\qquad\qquad\qquad (p_2^2+2y_{12}+2y_{23})^j (p_1^2+2y_{13})^l e^{-s p_3^2} \, \mathbbm d_1^2 \phi \, ,
\end{eqnarray}
and the last term,
\begin{eqnarray}
 \fl &\frac{1}{4} \int \rmd^dx \sqrt{\bar g} \int_0^\infty \rmd s \, \tilde f(s) \, \phi \sum_{j\geq0} \sum_{k\geq1} \frac{1}{(j+k+2)!} \left[ -s\bar\Delta, \left[ -s\mathbbm d_1, \left[ -s\bar \Delta, -s\mathbbm d_1 \right]_k \right] \right]_j e^{-s\bar\Delta} \phi \nonumber \\
 \fl &= \frac{1}{4} \int \rmd^dx \sqrt{\bar g} \int_0^\infty \rmd s \, \tilde f(s) \, \phi \sum_{j\geq0} \sum_{k\geq1} \frac{(-s)^{j+k+2}}{(j+k+2)!} (p_1^2+p_2^2+2y_{12}+2y_{13}+2y_{23})^j e^{-s p_3^2} \times \nonumber \\
 \fl &\qquad\qquad\qquad\qquad\qquad\qquad\qquad \bigg( \mathbbm d_1 \left[ \bar\Delta, \mathbbm d_1 \right]_k - \left[ \bar\Delta, \mathbbm d_1 \right]_k \mathbbm d_1 \bigg) \phi \nonumber \\
 \fl &= \frac{1}{4} \int \rmd^dx \sqrt{\bar g} \int_0^\infty \rmd s \, \tilde f(s) \, \phi \sum_{j\geq0} \sum_{k\geq1} \frac{(-s)^{j+k+2}}{(j+k+2)!} (p_1^2+p_2^2+2y_{12}+2y_{13}+2y_{23})^j e^{-s p_3^2} \times \nonumber \\
 \fl &\qquad\qquad\qquad\qquad\qquad\qquad\qquad \bigg( (p_1^2+2y_{13})^k - (p_1^2+2y_{12}+2y_{13})^k \bigg) \mathbbm d_1^2 \phi \, .
\end{eqnarray}
In both cases we thus need
\begin{eqnarray}
 \fl \phi \, \mathbbm d_1^2 \phi &= \phi \, \mathbbm d_1 \left[ \partial_3^\alpha \partial_3^\beta + \partial_1^\alpha \partial_3^\beta + \frac{1}{2} \bar g^{\alpha\beta} y_{13} \right] h_{\alpha\beta} \phi \nonumber \\
 \fl &= \left[ (\partial_1^\gamma + \partial_3^\gamma)(\partial_1^\delta + \partial_3^\delta) + \partial_2^\gamma (\partial_1^\delta + \partial_3^\delta) + \frac{1}{2} \bar g^{\gamma\delta} (y_{12} + y_{23}) \right] \times \nonumber \\
 \fl &\qquad\qquad\qquad\qquad\qquad\qquad \left[ \partial_3^\alpha \partial_3^\beta + \partial_1^\alpha \partial_3^\beta + \frac{1}{2} \bar g^{\alpha\beta} y_{13} \right] h_{\alpha\beta} h_{\gamma\delta} \phi \phi \nonumber \\
 \fl &= \bigg[ (3311) + (1311) + \frac{1}{2} y_{13} (\bar g 11) + 2(3313) + 2(1313) + y_{13} (\bar g 13) + (3333) \nonumber \\
 \fl &\qquad + (1333) + \frac{1}{2} y_{13} (\bar g 33) + (3312) + (1312) + \frac{1}{2} y_{13} (\bar g 12) \nonumber \\
 \fl &\qquad + (3323) + (1323) + \frac{1}{2} y_{13} (\bar g 23)+ \frac{1}{2}(y_{12}+y_{23}) (33 \bar g) \nonumber \\
 \fl &\qquad + \frac{1}{2}(y_{12} + y_{23}) (13 \bar g) + \frac{1}{4} y_{13}(y_{12} + y_{23}) (\bar g \bar g) \bigg] h_{\alpha\beta} h_{\gamma\delta} \phi \phi \, .
\end{eqnarray}
With the shorthand
\begin{eqnarray}
 \fl &\mathfrak X = \frac{1}{4} \int_0^\infty \rmd s \, \tilde f(s) \, \sum_{j\geq 0} \sum_{l\geq 0} \frac{(-s)^{j+l+2}}{(j+1)!(l+1)!} (p_2^2+2y_{12}+2y_{23})^j (p_1^2+2y_{13})^l e^{-s p_3^2} \nonumber \\
 \fl & \qquad +\frac{1}{4} \int_0^\infty \rmd s \, \tilde f(s) \, \sum_{j\geq0} \sum_{k\geq1} \frac{(-s)^{j+k+2}}{(j+k+2)!} (p_1^2+p_2^2+2y_{12}+2y_{13}+2y_{23})^j e^{-s p_3^2} \times \nonumber \\
 \fl & \qquad\qquad\qquad\qquad\qquad\qquad \bigg( (p_1^2+2y_{13})^k - (p_1^2+2y_{12}+2y_{13})^k \bigg) \, ,
\end{eqnarray}
we have the final contributions
\begin{eqnarray}
 \fl \mathfrak X &\addsto f^{(hh\phi\phi)}_{(3311)}, f^{(hh\phi\phi)}_{(1311)}, f^{(hh\phi\phi)}_{(3333)}, f^{(hh\phi\phi)}_{(1333)}, f^{(hh\phi\phi)}_{(3312)}, f^{(hh\phi\phi)}_{(1312)}, f^{(hh\phi\phi)}_{(3323)}, f^{(hh\phi\phi)}_{(1323)} \, , \nonumber \\
 \fl 2 \mathfrak X &\addsto f^{(hh\phi\phi)}_{(3313)}, f^{(hh\phi\phi)}_{(1313)} \, , \nonumber \\
 \fl \frac{1}{2} y_{13} \mathfrak X &\addsto f^{(hh\phi\phi)}_{(\bar g 11)}, f^{(hh\phi\phi)}_{(\bar g 33)}, f^{(hh\phi\phi)}_{(\bar g 12)}, f^{(hh\phi\phi)}_{(\bar g 23)} \, , \nonumber \\
 \fl y_{13} \mathfrak X &\addsto f^{(hh\phi\phi)}_{(\bar g 13)} \, , \nonumber \\
 \fl \frac{1}{2} (y_{12} + y_{23}) \mathfrak X &\addsto f^{(hh\phi\phi)}_{(33\bar g)}, f^{(hh\phi\phi)}_{(13\bar g)} \, , \nonumber \\
 \fl \frac{1}{4} y_{13} (y_{12} + y_{23}) \mathfrak X &\addsto f^{(hh\phi\phi)}_{(\bar g \bar g)} \, .
\end{eqnarray}
We can now compile the full list of all form factors that do not vanish in our ansatz:
\begin{eqnarray}
 \fl f^{(hh\phi\phi)}_{(\mathbbm 1)} = -\frac{1}{8} f(p_3^2) - \frac{1}{2} y_{13} \mathfrak Z \, , \\
 \fl f^{(hh\phi\phi)}_{(\bar g \bar g)} = \frac{1}{16} f(p_3^2) + \frac{1}{8} y_{13} \frac{f(p_1^2+2y_{13}+p_3^2)-f(p_3^2)}{p_1^2+2y_{13}} + \frac{1}{4} y_{13} (y_{12} + y_{23}) \mathfrak X \, , \\
 \fl f^{(hh\phi\phi)}_{(\bar g 11)} = f^{(hh\phi\phi)}_{(\bar g 12)} = f^{(hh\phi\phi)}_{(\bar g 23)} = f^{(hh\phi\phi)}_{(\bar g 33)} = \frac{1}{2} y_{13} \mathfrak X \, , \\
 \fl f^{(hh\phi\phi)}_{(\bar g 13)} = y_{13} \mathfrak X \, , \\
 \fl f^{(hh\phi\phi)}_{(13\bar g)} = f^{(hh\phi\phi)}_{(33\bar g)} = \frac{1}{4} \frac{f(p_1^2+2y_{13}+p_3^2)-f(p_3^2)}{p_1^2+2y_{13}} + \frac{1}{2} (y_{12} + y_{23}) \mathfrak X \, , \\
 \fl f^{(hh\phi\phi)}_{(23\bar g)} = \frac{1}{2} \mathfrak Z \, , \\
 \fl f^{(hh\phi\phi)}_{(1\bar g 3)} = f^{(hh\phi\phi)}_{(3\bar g 1)} = f^{(hh\phi\phi)}_{(3\bar g 3)} = -\mathfrak Z \, , \\
 \fl f^{(hh\phi\phi)}_{(1311)} = f^{(hh\phi\phi)}_{(1312)} = f^{(hh\phi\phi)}_{(1323)} = f^{(hh\phi\phi)}_{(1333)} = f^{(hh\phi\phi)}_{(3311)} = f^{(hh\phi\phi)}_{(3312)} = f^{(hh\phi\phi)}_{(3323)} = f^{(hh\phi\phi)}_{(3333)} = \mathfrak X , \\
 \fl f^{(hh\phi\phi)}_{(1313)} = f^{(hh\phi\phi)}_{(3313)} = 2\mathfrak X \, .
\end{eqnarray}

\section*{References}
\bibliographystyle{elsarticle-num}
\bibliography{general_bib}

%------------------------------------------------------------------
\end{document}